\documentclass[3p,twocolumn,authoryear]{elsarticle}

\usepackage{lineno}
\journal{Ocean Engineering}
\usepackage{graphicx}      % include this line if your document contains figures
\usepackage{amssymb}
\usepackage{amsmath}
\usepackage{color}
\usepackage[inline]{enumitem}
\usepackage{cuted}
\usepackage{flushend}
\usepackage{caption,subcaption}
\usepackage{nonfloat}
\usepackage{tabularx}

\def\ddt{\left(\frac{d}{dt}\right)}
\def\r{\mathbb{R}}

\def\vp{\varphi}
\def\re{\mathop{\rm Re}\nolimits}

\newdefinition{defn}{Definition}
\newtheorem{thm}{Theorem}
\newtheorem{rem}{Remark}

\begin{document}
\begin{frontmatter}

\title{Optimal Universal Controllers for Roll Stabilization}

%\thanks[footnoteinfo]{The work was supported in part by the European Research Council (ERCStG-307207), RFBR, grant 14-08-01015 and Russian Federation President's Grant MD-6325.2016.8.\\
%E-mail:\texttt{\{i.kapitaniuk,a.v.proskurnikov,m.cao\}@rug.nl}}

%Sponsor and financial support acknowledgment
%goes here. Paper titles should be written in uppercase and lowercase
%letters, not all uppercase.

\author[AGV,First]{Yuri A. Kapitanyuk\corref{corA}}
\ead{y.kapitanyuk@ieee.org}
\author[Second,Third]{Anton V. Proskurnikov}
\ead{anton.p.1982@ieee.org}
\author[First]{Ming Cao\fnref{footnoteinfo}}
\ead{m.cao@rug.nl}

\fntext[footnoteinfo]{
%The work was supported in part by 
M. Cao and Y. Kapitanyuk acknowledge support of the European Research Council (grant ERC-CoG-771687) and the Netherlands Organization for Scientific Research (grant NWO-vidi-14134). A. Proskurnikov acknowledges financial support of the RFBR (grant 20-01-00619).}
\cortext[corA]{Corresponding author}

\address[AGV]{Oceaneering AGV Systems B.V., Utrecht, The Netherlands}
\address[First]{Faculty of Science and Engineering, University of Groningen, The Netherlands}
\address[Second]{Department of Electronics and Telecommunications, Politecnico di Torino, Turin, Italy}
\address[Third]{Institute for Problems of Mechanical Engineering, St. Petersburg, Russia}
%\address[Navis]{Navis Engineering OY, R\&D Center, St. Petersburg, Russia}

\begin{abstract}
{Roll stabilization is an important problem of ship motion control. This problem becomes especially difficult if the same set of actuators (e.g. a single rudder) has to be used for roll stabilization and heading control of the vessel, so that the roll stabilizing system interferes with the ship autopilot. Finding the ``trade-off'' between the concurrent goals of accurate vessel steering and roll stabilization usually reduces to an optimization problem, which has to be solved in presence of an unknown wave disturbance. Standard approaches to this problem (loop-shaping, LQG, $H_{\infty}$-control etc.) require to know the spectral density of the disturbance, considered to be a ``colored noise''. In this paper, we propose a novel approach to optimal roll stabilization, approximating the disturbance by a \emph{polyharmonic} signal with known frequencies yet uncertain amplitudes and phase shifts. 
Linear quadratic optimization problems in presence of polyharmonic disturbances can be solved by means of the theory of \emph{universal} controllers developed by V.A. Yakubovich. An optimal universal controller delivers the optimal solution \emph{for any} uncertain amplitudes and phases. Using Marine Systems Simulator (MSS)
Toolbox that provides a realistic vessel's model, we compare our design method with classical approaches to optimal roll stabilization. Among three controllers providing the same quality of yaw steering, OUC stabilizes the roll motion most efficiently.
%and show that OUC controllers exhibit better performance.
}
\end{abstract}


\end{frontmatter}
%===============================================================================

\section{Introduction}

Roll stabilization is a classical problem in ship motion control~\citep{fossen1994guidance,perez2006ship,Perez2012129}.
%Excessive roll motion can cause discomfort for the crew and passengers, being also
%dangerous for fragile equipment and cargo.
Passive roll stabilization can be provided by special equipment such as
bilge keels, water-tanks and moving weights~\citep{perez2006ship,Perez2012129,rollTanks}; however,
these devices cannot be easily adapted to the unsteady environment and the changing wave's spectrum.
This limitation can be overcome by active (controlled) roll stabilization, which can be provided by
gyroscopic stabilizers, stabilizing fins and/or actuators (rudders and thrusters) used for the vessel's steering. This is illustrated by the \emph{rudder roll stabilization} (RRS), proposed originally for a vessels equipped with a single rudder~\citep{CowleyLambert1972,Carley1975,Lloyd1975}. Since fins, rudders and thrusters affect both
yaw and roll motion of the vessel, the roll stabilization controller should be integrated with the heading controller (autopilot).
These control systems can share some actuators and pursue concurrent goals of roll stabilization control and course steering.

A vessel's coupled yaw-roll motion can be modeled by a dynamical system, whose inputs are the rudder's and fins' angles and whose outputs stand for the ship's heading and roll. After linearizing this model, classical methods of linear control, e.g. loop shaping and Quantitative Feedback Theory~\citep{CowleyLambert1972,Carley1975,doi:10.1080/00207177808922376,blanke1993rudder,hearns98} can be applied to stabilize yaw and roll motion. To cope with nonlinearities, methods of feedback linearization and sliding mode control can be used~\citep{Lauvdal95nonlinearnon-minimum, GRIMBLE}.
For vessels equipped with fin stabilizers, classical methods usually decouple the roll motion from the yaw motion~\citep{FINS2007542,SOARES2015126}. However, ignoring internal cross-couplings often reduces the overall performance~\citep{ruddFinInterraction}.

The roll dynamics of a vessel appear to be non-minimum phase, leading thus to the \emph{fundamental limitation}~\citep{Carley1975,GoodwinPerez914671}: a controller stabilizing the vessel's heading cannot fully attenuate the wave-induced roll oscillations. A natural question arises, namely which level of the roll oscillation stabilization can be provided without deteriorating the yaw control. Mathematically, the latter goal is usually formulated as optimality of a special performance index, which penalizes the time-averaged steering error, roll angle and the control effort. Besides the control input, such a functional implicitly depends on the uncertain wave disturbance that affects the ship's motion.
Unlike the aforementioned stabilization techniques, optimization-based algorithms assume that some model of the disturbance is known. Most typically, the wave-induced motion is approximated by
either a ``colored noise'' or a random polyharmonic signal~\citep{Perez2012129,fossen1994guidance}.

The wave model of the first type approximates the wave disturbance by the output of some low-pass shaping filter, fed by a white noise. This approach, prevailing in the literature, reduces roll stabilization control design to standard methods of optimal controller synthesis, such as the
linear-quadratic Gaussian (LQG) control~\citep{van1987rudder,VANAMERONGEN1990679}, $\mathcal{H}_\infty$ control~\citep{Sharif1995703,blanke2000rudder,Crossland2003423,381212} and model-predictive control (MPC)~\citep{perez2006ship}. As usual in stochastic and minimax control, optimal controllers do not deliver optimal solutions for any specific realization of the stochastic disturbance, providing optimality either ``on average'' (in the sense of expectation) or in the ``worst-case'' scenario. Another downside of the mentioned methods is the necessity to estimate the spectral density of the wave motion.

An alternative ``discrete'' model of the wave motion, often used in marine engineering~\citep{perez2006ship,Nicolau2005,irregular_waves}, approximates the  wave motion by
the sum of sinusoids with known frequencies, where the constant amplitudes are obtained via sampling of the spectral density and random phase shifts are uniformly distributed in $[0,2\pi]$ in order to get different realizations. For this model of the wave disturbance and linearized vessel's yaw-roll dynamics, the optimal roll stabilization may be considered as a linear-quadratic optimization problem, where the control system is affected by a partially uncertain polyharmonic signal. A relevant extension of the classical LQR control to cope with such problems has been developed in~\citep{Yakubovich1995,587333,788535,Proskurnikov2006,Proskurnikov2012,Proskurnikov2015557}. It appears that (under natural assumptions) an \emph{optimal universal controller} (OUC) exists, which is independent of the uncertain signal's parameters, delivers the optimal
process for \emph{arbitrary} values of these parameters. Furthermore, the OUC can be found in the class of linear stabilizing controllers; a convenient parametrization of such OUCs has been found~\citep{Yakubovich1995}. 

In this paper, we apply Yakubovich's theory  of OUC to the problem of optimal roll stabilization. This paper extends our previous work~\citep{cams}, which considered a simplified model of the vessel with a single rudder and no stabilizing fins. We illustrate the efficiency of OUCs in the optimal roll stabilization problem and compare it with classical controllers by using numerical simulations that utilize the ``benchmark'' vessel's model from~\citep{perez2006ship}. The OUC theory provides a method for combined fin-rudder stabilization control design, avoiding the undesired counteraction between different actuators and improving the resulting efficiency of the control system. Unlike the usual LQR~\citep{Perez2012129}, the OUC does need to measure the full state vector and provides optimality for any polyharmonic signal from the specified class; to find OUC, one does not need to solve the Riccati equation. Unlike LQG and $H_{\infty}$ approaches, the OUC design does not require one to know the spectral density of the wave motion (or, equivalently, the structure of the shaping filter). The OUC depends only on the fixed wave's frequencies and ensures optimality of the cost functional \emph{for any} realization of the random disturbance.

The paper is organized as follows.
In Section~\ref{sec:model}, mathematical models of the vessel's motion and wave disturbances are considered.
In Section~\ref{sec:theory} the theory of OUC in general problems of linear-quadratic optimization with uncertain disturbances is introduced. In Section~\ref{sec:rrd}, we apply this theory to design an optimal roll stabilization controller, whose performance is studied numerically in Section~\ref{sec:sim}.

\section{Mathematical models}\label{sec:model}

We first introduce mathematical models of the ship's yaw-roll motion and the wave disturbances.

\subsection{The vessel's motion}

The movements of a marine vessel (as a rigid body) have six degrees of freedom. The standard 6-DoF mathematical model can be found in~\citep{Perez2012129,fossen1994guidance}. However, it is more convenient to use a simplified reduced-order model~\citep{VANAMERONGEN1990679,fossen1994guidance,perez2006ship}, which is derived {(see details in~\ref{app:4dof})} under two simplifying assumptions: 1)the effects of the pitch and heave motion of the vessel on its surge, sway, roll and yaw dynamics are negligibile; 2) the vessel's speed is changing slowly relative to the remaining coordinates. Under these assumptions, the yaw and the roll controllers can be designed for a simplified linearized model.

In the original papers on rudder roll stabilization~\citep{CowleyLambert1972,Lloyd1975},
the simplest configuration of the vessel with one rudder has been considered, whose angle is the single control input of the system. In general, the vessel can be equipped with multiple actuators (rudders, azimuth and tunnel thrusters, waterjets etc.); however, for the sake of autopilot and roll stabilization control design they are usually replaced by an equivalent ``virtual rudder'', whose ``angle'' stands for the scaled rotating yaw moment, distributed among the actuators by a separate \emph{control allocation} system~\citep{Johansen2008}. In addition to this, we allow the vessel to have synchronized stabilizing fins, whose angle serves as the second control.

\begin{figure}
	\begin{center}
		\includegraphics[width=0.45\columnwidth]{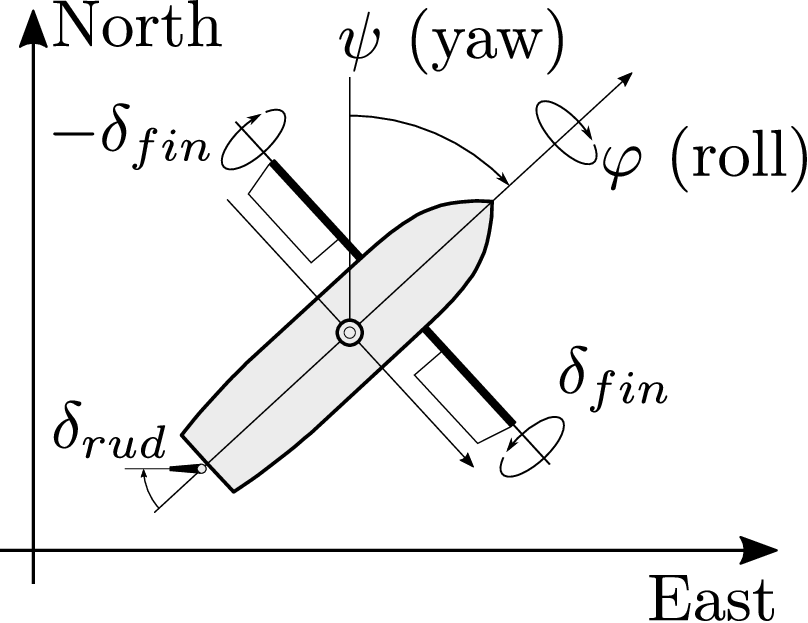}    % The printed column width is 8.4 cm.
		\caption{ The rudder ($\delta_{rud}$), the fin ($\delta_{fin}$),  roll ($\vp$) and yaw ($\psi$) angles.}
		\label{fig:vessel0}
	\end{center}
\end{figure}

\begin{figure}
	\begin{center}
		\includegraphics[width=0.6\columnwidth]{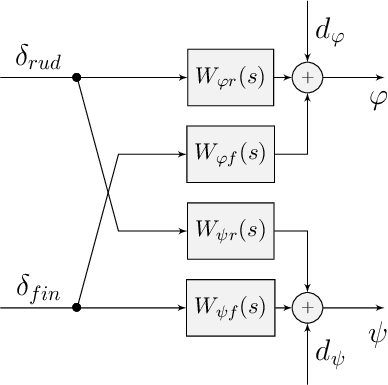}
		\caption{The reduced model of vessel's dynamics}
		\label{fig:mod}
	\end{center}
\end{figure}

Denoting the rudder, the fin, the roll and the yaw (or heading) angles by, respectively, $\delta_{rud}(t)$, $\delta_{fin}(t)$, $\vp(t)$ and $\psi(t)$ (Fig.~\ref{fig:vessel0}), the reduced-order vessel's model
has the structure illustrated in Fig.~\ref{fig:mod}. The system is affected by the environmental disturbance, represented by its roll and yaw components\footnote{For clarity, in this paper we consider the ``motion superposition'' model~\citep{perez2006ship}, where the disturbance is modeled as an uncertain displacement from the original trajectory of the vessel. An alternative approach, referred to as the ``force superposition''~\citep{perez2006ship}, treats the disturbance as an additional force, acting on the ship's hull.} $d_{\vp}(t)$, $d_{\psi}(t)$.
The transfer functions from $\delta_{rud}$ and $\delta_{fin}$ to $\vp$ and $\psi$, denoted by $W_{\vp r}(s)$, $W_{\vp f}(s)$, and $W_{\psi r}(s)$, $W_{\psi r}(s)$ respectively, are as follows~\citep[Sect.~8.2]{perez2006ship}
\begin{equation}\label{eq:vessel_model2}
\begin{gathered}
W_{\vp r}(s) = \frac{K_{\vp r}(q_1-s)(q_2+s)}{(p_1+s)(p_2+s)(s^2 + 2 \zeta_{\vp}\omega_{\vp} s + \omega^2_{\vp})},\\
W_{\psi r}(s) = \frac{K_{\psi r} (q_3+s)(s^2 + 2 \zeta_{q}\omega_{q} s + \omega^2_{q})}{s(p_1+s)(p_2+s)(s^2 + 2 \zeta_{\vp}\omega_{\vp} s + \omega^2_{\vp})},\\
W_{\vp f}(s) = \frac{K_{\vp f}(q_4-s)(q_5+s)}{(p_1+s)(p_2+s)(s^2 + 2 \zeta_{\vp}\omega_{\vp} s + \omega^2_{\vp})},\\
W_{\psi f}(s) = \frac{K_{\psi f} (q_6-s)(s^2 + 2 \zeta_{t}\omega_{t} s + \omega^2_{t})}{s(p_1+s)(p_2+s)(s^2 + 2 \zeta_{\vp}\omega_{\vp} s + \omega^2_{\vp})},
\end{gathered}
\end{equation}
where $q_i>0$, $p_j>0$, $\omega_{\vp},\omega_q, \omega_t>0$ and $\zeta_{\vp},\zeta_q,\zeta_t\in (0;1)$ are constants.
{Notice that this model takes into account coupling between the sway, roll and yaw motions of the vessel. Ignoring the cross-coupling between yaw and roll, the model can be further reduced~\cite[Section 9.1.1]{fossen1994guidance}.}

Along with the transfer function, one can introduce the state-space model of the system 
\begin{equation}\label{eq:vessel_model}
\begin{aligned}
\dot{x}(t) &= A x(t) + B \delta(t)\\
y(t) &= C x(t) + G d(t).\\
\end{aligned}
\end{equation}

Here the vessel's reduced state vector $x(t)=(\vp,p,\psi,r,v)^{\top}$ consist of the roll angle $\vp$, the roll rate $p=\dot\vp$, the yaw angle $\psi$, the yaw rate $r=\dot\psi$ and the sway velocity $v$. The disturbance $d(t) = (d_\varphi,d_\psi)^{\top}$ stands for the wave-induced motion of the vessel. The vector $y(t) = (\varphi,\psi)^{\top}\in \mathbb{R}^2$ stands for the
system's output, whose components $\varphi$ and $\psi$ are measured, respectively, by a vertical reference unit (VRU) sensor~\citep{Balloch1998} and a gyro or GPS compass and the control input is presented by the vector $\delta(t) = (\delta_{rud},\delta_{fin})^{\top}$.

{The explicit derivation of the matrices $A,B,C,G$ is given in~\ref{app:4dof}. It should be noticed that the controller design, in fact, does not use their explicit values and requires only the knowledge of the transfer functions~\eqref{eq:vessel_model2}.}

% \textcolor{red}{Explain that the linearization and controller based on it are applicable for the fixed longitudinal speed. When the speed is changed, the model and the controller have to be re-initialized. This corresponds to the procedure of gain scheduling (Khalil). Answering the question of reviewer 2, refer to this remark.}

\subsection{The disturbance model}

The environmental disturbances, influencing a marine craft's motion, are due to the waves, the wind and the current.
The fast oscillations in the roll and the heading angles are mainly caused by the waves,
whereas the current and the wind are changing much more slowly and their effect is usually modeled as a constant roll angle and stationary heading deviation.
Henceforth, the disturbance $d(t)$ stands for the wave-induced motion only. In this paper, we use a \emph{polyharmonic} approximation of
this motion~\citep{Perez2012129,fossen1994guidance}
\begin{equation}\label{eq:wave11}
\begin{gathered}
d_{\vp}(t) = \sum_{i = 1}^{p} a_i^{\vp}   \sin \left( \omega_i t +\phi_i^{\vp}\right),\\
d_{\psi}(t) = \sum_{i = 1}^{p} a_i^{\psi} \sin \left( \omega_i t +\phi_i^{\psi}\right).
\end{gathered}
\end{equation}
Here the spectrum $\omega_1,\ldots,\omega_p \ge 0$ is known.
The special case $p =1$ corresponds to the model of \emph{regular} waves; however, a real state of the sea is best described by a random or \emph{irregular} wave model.
This stochastic process can be approximated by the model \eqref{eq:wave11} with $p$ being sufficiently large. The constant amplitudes $a_i^{\vp}$ and $a_i^{\vp}$ are obtained via sampling the spectral density with a small enough step $\Delta \omega$ to ensure that the fundamental period of the finite sum of sinusoidal components is longer than the desired duration of the simulation. The random phase shifts $\phi_i^{\vp}$ and $\phi_j^{\vp}$ used to generate different realizations of the stochastic process are uniformly distributed in $[0,2\pi]$. Although the model~\eqref{eq:wave11} of irregular waves can describe a sea state quite accurately, the direct use of it in the control design is difficult due to the high dimension. The better strategy is to consider a few ``dominating'' frequencies corresponding to the peaks of the spectral density. In general, the localization and the shape of the spectral density highly depend on many parameters of motion such as the average speed of the vessel, sailing conditions and a frequency response of the vessel's hull; however, these ``dominating'' frequencies
can be efficiently estimated in real time, see e.g.~\citep{Belleter201548,6145613,Fedele2012,Hou2012} and references therein. For simplicity and clarity of presentation, we proceed to assume that the number and the values of such frequencies are \emph{known}.

It should be noted that in the existing control literature the wave motion is usually approximated by the ``colored noise'', that is, the output from a low-pass shaping filter fed by the white noise signal. The simplest approximation for the shaping filter's transfer function (that is, the wave spectrum), is
\begin{equation}\label{eq:wave1}
H(s) = \frac{K_w s}{s^2+2 \zeta_0 \omega_0 s + \omega_0^2}.
\end{equation}
Here the constant $K_w>0$ determines the wave strength, $\omega_0$ is the encounter frequency and $\zeta_0$ is the damping ratio~\citep{Perez2012129}.
Unlike our approach, using only the information about the frequencies, the existing approaches, as discussed in Introduction, typically use all parameters of the transfer function $H(s)$, whose
identification is a self-standing non-trivial problem.
{As discussed in~\cite[Sect. 2.5]{Perez2012129}, the shaping filter representation of the wave is primarily used in stochastic control, which is convenient for methods exploiting spectral factorization of the wave disturbance, while the more precise nolinear multi-sine model is commonly used in naval architecture.}

\section{Linear-quadratic optimization in presence of uncertain polyharmonic signals}\label{sec:theory}

In this section, the basic ideas of the theory of OUC are given for the reader's convenience, following the survey paper~\citep{Proskurnikov2015557}. The concept of universal controller dates back to early works on ``signal invariance'', or disturbance decoupling in control systems, see e.g. the survey in~\cite{ProYak:03b,ProYak:03a}.

We start with introducing some notation. The set of complex $m\times n$ matrices is denoted by $\mathbb{C}^{m\times n}$. The Hermitian complex-conjugate transpose of a matrix $M\in\mathbb{C}^{m\times n}$ is denoted by $M^*\in\mathbb{C}^{n\times m}$. We use $\imath \triangleq \sqrt{-1}$ to denote the imaginary unit.
The real part of a number $z\in \mathbb{C}$ is denoted by $\re z$.
%Given a quadratic form $\mathcal F(x)=x^{\top}Fx$ (where $F=F^{\top}$ is a symmetric real matrix), its Hermitian prolongation to the complex domain is denoted by $\tilde{\mathcal F}(\tilde x)=\tilde %x^*F\tilde x$.

\subsection{A family of uncertain optimization problems}

Consider a linear time-invariant MIMO system, influenced by an exogenous signal
\begin{align}\label{eq:1}
\dot{x}(t) &= A x(t) + B u(t) + E d(t), \nonumber \\
y(t) &= C x(t) + D u(t) + G d(t).
\end{align}

Here $x(t) \in \mathbb{R}^n, u(t) \in \mathbb{R}^m, y(t) \in \mathbb{R}^k$
stand for, respectively, the state vector, the control and the observed output.
The signal $d(t) \in \mathbb{R}^l$ is a polyharmonic process with \emph{known} spectrum $\omega_1,\ldots,\omega_N$
\begin{equation}\label{eq:2}
d(t) = \re\sum_{j=1}^{N}{d_j e^{\imath \omega_j t}},
\end{equation}
whose complex amplitudes $d_i\in\mathbb{C}^l$ (absorbing also the phase shifts) are uncertain.
The components of this exogenous signal may include disturbances, measurement noises and reference signals.

In presence of the oscillatory disturbance~\eqref{eq:2}, the solutions of~\eqref{eq:1} do not vanish at infinity.
The goal of control is to guarantee boundedness of the solution $(x(t),u(t))$ and its optimality in the sense of the following quadratic performance index
\begin{equation}\label{eq:3}
J\left[ x,u,d\right]= \varlimsup_{T \rightarrow \infty} \frac{1}{T} \int_0^{T}
\mathcal{F} \left[ x(t), u(t), d(t)\right] dt.
\end{equation}
Here $\mathcal{F}$ is a quadratic form, which is assumed to be non-negative definite $\mathcal F\ge 0$. Considering the integrand in~\eqref{eq:3} as a measure of the solution's ``energy'',
its average value $J$ can be thought of as the solution's average ``power''. Formally, the control goal can be formulated as follows
\begin{equation}\label{eq:4}
\begin{gathered}
\text{minimize $J(x(\cdot),u(\cdot),d(\cdot))$}\\
\text{ subject to (\ref{eq:1}) and } \sup_{t \geq 0} (|x(t)|+ |u(t)|) < \infty.
\end{gathered}
\end{equation}

In fact,~\eqref{eq:4} defines an infinite family of optimization problems, corresponding to different choices of the amplitudes $d_1,\ldots,d_N$. Obviously, the set of optimal processes also depends on the amplitudes and hence cannot be found explicitly. Nevertheless, it can be shown that an \emph{optimal universal} controller (OUC) exists that provides an optimal process \emph{for any} uncertain amplitudes $d_i$, solving thus the whole family of optimization problems~\eqref{eq:4}.
\begin{defn}
A causal operator $\mathcal U:y(\cdot)\mapsto u(\cdot)$ is an OUC for the family of optimization problems~\eqref{eq:4}, if for any initial condition $x(0)\in\r^n$ and any amplitudes $d_1,\ldots,d_N$ in~\eqref{eq:2} there exists a unique solution of the closed-loop system
\[
\dot{x} = A x + B u + E d,\,y=Cx+Du+Gd,\quad u(\cdot)=\mathcal Uy(\cdot),
\]
which is bounded and delivers an optimum to~\eqref{eq:4}.
\end{defn}

\subsection{A class of linear OUC}

Although the existence of OUCs may seem exceptional, such controllers exist under rather mild assumptions on the system and the cost functional.

We assume that the system~\eqref{eq:1} is stable, that is, $\det(sI_n-A)\ne 0$ whenever $\re s\ge 0$. If the system is stabilizable and detectable, one may always augment it
with an observer-based stabilizing controller, so the stability assumption can be adopted without loss of generality.

%\begin{assum}
Let $ F= F^{\top}$ stand for the matrix of the quadratic form $\mathcal{F}(x,u,d)$ and $ F_0= F^{\top}_0$ be the matrix of the quadratic form $\mathcal{F}_0(x,u)=\mathcal{F}(x,u,0)$, that is,
\begin{equation}\label{eq:5}
\begin{aligned}
\mathcal{F}(x,u,d)&=
\left[
\begin{smallmatrix}
	x \\ u \\ d
\end{smallmatrix}
\right]^{\top}
 F
\left[
\begin{smallmatrix}
	x \\ u \\ d
\end{smallmatrix}
\right]=\left[\begin{matrix}
	x \\ u
\end{matrix}
\right]^{\top}
 F_0
\left[
\begin{matrix}
	x \\ u
\end{matrix}
\right]+
\\
&+ 2d^{\top} F_{dx} x + 2d^{\top}F_{du}u + d^{\top}F_{dd}d,
\end{aligned}
\end{equation}
where $F_{dx},F_{du},F_{dd}=F_{dd}^{\top}$ are matrices of appropriate dimensions. We introduce the rational complex-valued matrix $\Pi(\imath\omega)=\Pi(\imath\omega)^*$ as follows
\begin{equation*}
\tilde{u}^* \Pi(\imath \omega) \tilde{u} =
\begin{bmatrix}
	A_{\imath\omega}^{-1}B\tilde u \\ \tilde u
\end{bmatrix}
^*{F}_0
\begin{bmatrix}
	A_{\imath\omega}^{-1}B\tilde u \\ \tilde u
\end{bmatrix}
,\, A_s:=sI_n-A,
\end{equation*}
and assume that the \emph{frequency-domain} condition holds
\begin{equation}\label{eq:6}
\Pi(\imath\omega)\ge\varepsilon I_m,\quad\varepsilon=const>0.
\end{equation}
The condition (\ref{eq:6}) is a standard solvability condition for classical LQR problems, providing the existence of the stabilizing solution to the Riccati equation~\citep{Anderson:1990:OCL:79089}.
It always holds when $F_0(x,u)$ is positively definite, which is a natural assumption in practice.
The condition(\ref{eq:6}) cannot be discarded and, moreover, its ``strong'' violation in the sense that $\tilde{u}^* \Pi(\imath \omega_0) \tilde{u} < 0$ for some $\omega_0 \in \mathbb{R}$ and $\tilde{u} \in \mathbb{C}^m$ implies\footnote{For a similar discrete-time optimization problem, the proof is available in~\citep{788535}, and the continuous-time case is considered in the same way.} the ill-posedness~of the problem~\eqref{eq:4}: $\inf J=- \infty$ for any signal~\eqref{eq:2}.

Under non-restrictive assumptions, the OUC exists and can be found among linear controllers
\begin{equation}\label{eq:9}
N \ddt u(t) = M \ddt y(t),
\end{equation}
where $N$ and $M$ stand for matrix polynomials; the matrix $N(s)$ is square and $\det N\not\equiv 0$.
The relevant result is given by the following theorem.
\begin{thm}\citep{Proskurnikov2015557}~\label{thm:1}
Let the system~\eqref{eq:1} be stable and the inequality~\eqref{eq:6} hold. Then the linear controller~\eqref{eq:9} is an OUC for the family of problems~\eqref{eq:4} if the following two conditions hold
\begin{enumerate}
\item the closed-loop systems is stable, that is,
\begin{equation}\label{eq:8}
\det 
\begin{bmatrix}
s I_n - A & -B \\
- M(s)C & N(s) - M(s)D
\end{bmatrix}
\neq 0,
\end{equation}
\[
\forall s: \re  s \geq 0;
\]
\item the \textbf{closed-loop} transfer function $W_{ud}$ from $d$ to $u$ satisfies the interpolation equations
\begin{equation}\label{eq:10}
W_{ud}(\imath \omega_j) = R_j, \ \forall j = 1,2,...,N,
\end{equation}
where the constant matrices $R_j$ are as follows
\[
R_j = - \Pi^{-1}(\imath \omega_j)
\left[
\begin{matrix}
	A_{\imath \omega_j}^{-1} B\\
	I_m\\
	0
\end{matrix}
\right]^*
{F}
\left[
\begin{matrix}
	A_{\imath \omega_j}^{-1} E\\
	0\\
	I_l
\end{matrix}
\right].
\]
\end{enumerate}
\end{thm}

Note that, unlike the classical LQR problem, where the optimal controller is uniquely defined from the Riccati equation, the OUC in the problem~\eqref{eq:4} is not unique; to find it, one need not solve Riccati equations. We will use Theorem~\ref{thm:1} in a special situation, where $\mathcal F$ depends only on the output and the control,
i.e. $F$ admits the decomposition
\begin{equation}\label{eq:f-hat}
F=\begin{bmatrix}
C & D & G\\
0 & I_m & 0
\end{bmatrix}^*\hat F\begin{bmatrix}
C & D & G\\
0 & I_m & 0
\end{bmatrix},
\end{equation}
where $\hat F=\hat F^*\in\mathbb{C}^{m+n}$. In this situation, one has
\begin{equation}\label{eq:tf-special}
\begin{gathered}
\Pi(\imath\omega)=\begin{bmatrix}
W_{yu}^0(\imath\omega)\\
I_m
\end{bmatrix}^*\hat F\begin{bmatrix}
W_{yu}^0(\imath\omega)\\
I_m
\end{bmatrix},\\
R_j = - \Pi^{-1}(\imath \omega_j)\begin{bmatrix}
W_{yu}^0(\imath\omega_j)\\
I_m
\end{bmatrix}^*\hat F\begin{bmatrix}
W_{yd}^0(\imath\omega_j)\\
0
\end{bmatrix}.
\end{gathered}
\end{equation}
Here $W_{yu}^0(s)$ and $W_{yd}^0(s)$ stand for the \textbf{open-loop} transfer functions from respectively $u$ and $d$ to $y$
\[
%\begin{gathered}
W^0_{yu}(s):=CA_s^{-1}B + D,\;W^0_{yd}(s):=CA_s^{-1}E + G.
%\end{gathered}
\]

Recalling that $A$ is a Hurwitz matrix, it can be shown that the closed-loop system is stabilized by the controller~\eqref{eq:9}, whose coefficients are as follows
\begin{equation}\label{eq:13}
\begin{gathered}
M(s) = \Delta(s) r(s),\\N(s) = M(s) \left[CA_s^{-1}B + D\right] + \rho(s) I_m,\\
\Delta(s) := \det(A_s)=\det(sI_n-A).
\end{gathered}
\end{equation}
Here $r(s)$ is a matrix polynomial and $\rho(s)$ is a \emph{scalar} Hurwitz polynomial with $\deg\rho\ge\deg M$.
Such a controller is ``feasible'' in the sense that its transfer matrix $N^{-1}M$, as well as the closed-loop system's transfer matrices from $d$ to $x,u$, are proper.
For the controller~\eqref{eq:9},\eqref{eq:13}, one obtains
\begin{equation}\label{eq:14}
W_{ud}(s)= \frac{M(s)}{\rho(s)}W^0_{yd}(s),
\end{equation}
and the interpolation constraints~(\ref{eq:10}) boil down to
\begin{equation}\label{eq:15}
\Delta(\imath \omega_j) r(\imath \omega_j) W^0_{yd}(\imath\omega_j) = \rho (\imath \omega_j) R_j.
\end{equation}

The constraints~\eqref{eq:15} can be satisfied when
\begin{equation}\label{eq:16}
\begin{gathered}
%\rk W^0_{yd}(\imath\omega_j)= l = \dim d(t),\quad \forall j\\
\det\left[W^0_{yd}(\imath\omega_j)W^0_{yd}(\imath\omega_j)^*\right]\ne 0\;\forall j=1,\ldots,N.
\end{gathered}
\end{equation}
Here $W_{yd}^0$ is the open-loop transfer matrix from $d$ to $y$. The conditions~\eqref{eq:16} typically hold when $\dim y\ge\dim d$.
Furthermore, if~\eqref{eq:16} holds, the coefficients of $r$ and $\rho$ can be chosen as continuous functions of $\omega_j$, so that the controller is robust to small deviations in the spectrum
$\omega_j^{'} \approx \omega_j$. Choosing an arbitrary Hurwitz polynomial $\rho$ of degree $\deg\rho \geq 2N+\deg\delta-1$, one needs to find the matrix polynomial $r$ with $\deg r\le 2N-1$, satisfying the conditions
\begin{equation}\label{eq:15a}
%\frac{r(\imath\omega_j)}{\rho (\imath \omega_j)}
\begin{gathered}
r(\imath\omega_j)=r^0(\imath\omega_j),\\
r^0(s):=\frac{\rho (s)R_j}{\Delta(s)}W^0_{yd}(s)^*\left[W^0_{yd}(s)W^0_{yd}(s)^*\right]^{-1}.
\end{gathered}
\end{equation}
Separating the real and imaginary parts, one obtains $2N$ equations for $2N$ real coefficients of $r$.

It appears that any OUC~\eqref{eq:9} is equivalent, in some sense~\citep{Yakubovich1995,Proskurnikov2015557}, to the controller~\eqref{eq:13} with some polynomials $r,\rho$, satisfying the interpolation constraints~\eqref{eq:15}.

\begin{figure}[h]
	\begin{center}
		\includegraphics[width=0.75\columnwidth]{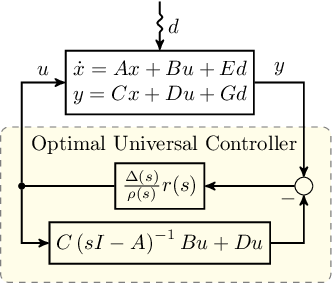}    % The printed column width is 8.4 cm.
		\caption{The structure of the OUC~\eqref{eq:13}}
		\label{fig:scheme}
	\end{center}
\end{figure}

\begin{rem}\label{rem:tf}
Note that the controller~\eqref{eq:13} in fact does not depend on the state-space model~\eqref{eq:1}, involving only the system's characteristic polynomial
$\Delta(s)$ and the open-loop transfer function $W_{yu}^0(s):=D+C(sI-A)^{-1}B$ from $u$ to $y$ (Fig.~\ref{fig:scheme}).
In the case where $\mathcal F=\mathcal F(y,u)$ depends only on $y$ and $u$,
the interpolation conditions~\eqref{eq:15} also involve only the values of $W_{yu}^0(\imath\omega_j)$ and $W_{yd}^0(\imath\omega_j)$ rather than the whole state model~\eqref{eq:1}. Hence, in this special situation, the design of OUC requires only the knowledge of $\Delta(s)$, $W_{yu}^0(s)$ and $W_{yd}^0(s)$, which are independent of the minimal state-space realization.
\end{rem}

{\begin{rem}\label{rem.bandpass}
In general, one has a lot of freedom in choosing the coefficients of $\rho(s)$, and their ``optimal'' choice remains an important research topic as a subject of ongoing research. In practice, the polyharmonic model of the disturbance is usually imprecise: the signal $d(t)$ contains frequencies other than $\omega_j$ (whose amplitudes are sufficiently small). Fig.~\ref{fig:scheme} suggests that, ideally, the OUC's transfer function $r(s)\Delta(s)/\rho(s)$ should have a sufficiently narrow bandpass, containing the frequencies $\omega_j$ in order to damp these unmodeled spectrum. Although this requirement is not very formal, it can be used for practical tuning of the OUC's parameters.
\end{rem}}

\begin{rem}
As discussed in~\citep{788535}, the important property of the OUC~\eqref{eq:9} is its \emph{robustness} against small changes in the frequencies $\omega_j$, whereas the straightforward LQR-based design leads to a controller that is formally optimal yet non-robust to deviations in spectrum. The results from~\citep{788535} deal with discrete-time systems, but this robustness property is retained by the continuous-time OUC~\eqref{eq:9}.
\end{rem}

\section{Optimal Universal Roll Stabilization Controllers}\label{sec:rrd}

In this section, we reduce the optimal roll stabilization problem to a special case of the problem~\eqref{eq:4}.
The cost functional will depend only on the control effort and output.
In view of Remark~\ref{rem:tf}, in this situation one does not need to know a special state-space representation of the open-loop system, requiring only its characteristic polynomial and transfer matrices $W_{yu}^0,W_{yd}^0$. In this sense, an optimal controller can be designed in the frequency domain.

\begin{figure}[h]
	\begin{center}
		\includegraphics[width=0.75\columnwidth]{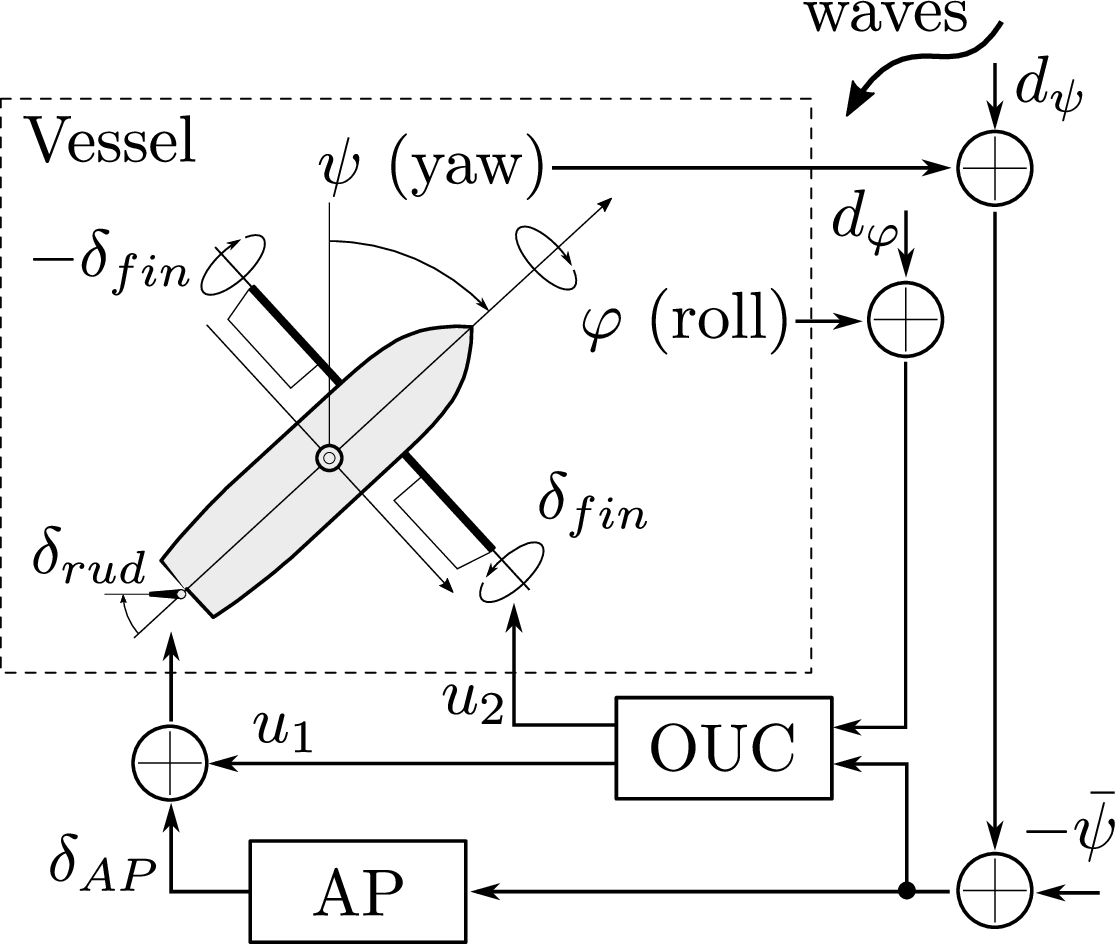}    % The printed column width is 8.4 cm.
		\caption {The vessel's control system: the autopilot (AP) and the optimal universal controller (OUC)  for roll stabilization.}
		\label{fig:vessel}
	\end{center}
\end{figure}

We assume that the vessel's heading is stabilized by a known \emph{autopilot} (Fig.~\ref{fig:vessel}).
Behind this statement, there are two practical considerations. First of all, it allows splitting of the adjustment procedure for a motion control system on the vessel in two sequential stages:  the independent tuning of an autopilot and the following design of the roll stabilization controller. The second reason is the flexibility and the modularity; the roll stabilization system may be supplied by a manufacturer of the equipment such as high-performance rudders or active fins independent of the development of the autopilot, which is in itself a challenging task.
 The autopilot design problem has been thoroughly studied in the literature~\citep{fossen1994guidance,perez2006ship,Nicolau2005,Veremey2014} and is beyond the scope of this paper.
Furthermore, we assume that the roll stabilization system is aware of the measured heading of the vessel and the constant heading setpoint $\bar\psi$.
	In practice, $\bar\psi(t)$ can be a function of time, e.g. when autopilot steers the vessel along a curvilinear path. However, these dynamics are much slower than the ship's roll motion, and hence are neglected in the roll stabilization system design.
The deviation among them (heading error) $e_{\psi}(t)$, along with the roll stabilization error $e_{\varphi}(t)$ are the inputs to the roll stabilization system (Fig.~\ref{fig:vessel}). Mathematically,
\begin{equation*}\label{eq:rel1}
e_{\psi}(t) := \psi(t) + d_{\psi}(t)- \bar{\psi},\quad e_{\varphi}(t) := \varphi(t) + d_{\varphi}(t).
\end{equation*}

The rudder angle $\delta_{rud}(t)$ is the sum of the autopilot's and the roll stabilization  controller's commands (Fig.~\ref{fig:vessel}), denoted respectively by $\delta_{AP}(t)$ and $u_1(t)$. The fin angle $\delta_{fin}(t)$ is used as the second  control input $u_2(t)$.
Denoting the autopilot's transfer function by $W_{AP}(s)$, one has
\begin{equation*}\label{eq:rel2}
\begin{aligned}
\delta_{rud}(t) &= \delta_{AP}(t)+u_1(t)=W_{AP}\ddt e_\psi (t) + u_1(t)\\
\delta_{fin}(t)&=u_2(t).
\end{aligned}
\end{equation*}

The yaw-roll dynamics of the vessel, closed by the autopilot, are represented by the input-output model
\begin{equation}\label{eq:rel23}
\begin{gathered}
y(t)=W_{yu}^0\ddt u(t)+W_{yd}^0 \ddt d(t),\\
y(t):=
\begin{bmatrix}
e_\varphi(t)\\
e_\psi(t)
\end{bmatrix}, \
u(t):=
\begin{bmatrix}
u_1(t)\\
u_2(t)
\end{bmatrix}, \
d(t):=\left[\begin{matrix}
\bar{\psi}\\ d_\varphi(t)\\ d_\psi(t)
\end{matrix}\right].
\end{gathered}
\end{equation}
Here $d_{\vp}(t),d_{\psi}(t)$ are the polyharmonic components of the wave-induced motion~\eqref{eq:wave11}.
Considering $\bar\psi$ as a harmonic signal of zero frequency, $d(t)$ is a special case of~\eqref{eq:2} with $l=3$ and $N=1+p$, where $\omega_j$, $k=1,\ldots,p$ are the wave frequencies from~\eqref{eq:2} and $\omega_{1+p}=0$. The transfer functions $W_{yu}^0,W_{yd}^0$ depend on the autopilot's transfer function $W_{AP}$ (from $e_{\psi}$ to $\delta_{AP}$) and the functions $W_{yaw},W_{roll}$ from~\eqref{eq:vessel_model2}. The exact formulas for $W_{yu}^0,W_{yd}^0$ are derived in~\ref{app:Model} and it can be easily seen from these formulas that~\eqref{eq:16} always holds for any wave $\omega_1,\ldots,\omega_N\in\r$.

The cost functional penalizes the mean square values of the following three variables
\begin{enumerate*}[label=(\roman*)]
	\item the roll displacement ($e_{\phi}$),
	\item the heading deviation ($e_{\psi}$), and
	\item the control effort
\end{enumerate*}.
Denoting the corresponding penalty weights by $\alpha,\beta,\gamma_{1,2}>0$, we introduce the quadratic cost functional  as follows
\begin{equation}\label{eq:J1}
\begin{gathered}
J= \varlimsup_{T \rightarrow \infty} \frac{1}{T} \int_0^{T}\mathcal F(y(t),u(t))\,dt,\\
%\left(\alpha e_\varphi(t)^2 + \beta e_\psi(t)^2 + \gamma  u(t)^2\right)dt.
\mathcal F(y,u):=\alpha e_\varphi^2 + \beta e_\psi^2 + \gamma_1  u_1^2+\gamma_2u_2^2.
\end{gathered}
\end{equation}
The Hermitian form $\mathcal F$ can be represented in the form~\eqref{eq:f-hat}, where $\hat F$ is defined by
\[
\hat F=
\begin{pmatrix}
\alpha & 0 & 0 & 0 \\ 0 & \beta & 0 & 0 \\ 0 & 0 & \gamma_1 & 0\\0 & 0 & 0 & \gamma_2
\end{pmatrix}.
\]
The matrix function $\Pi(\imath\omega)$ and the matrices $R_j$ are defined by~\eqref{eq:tf-special}; $\Pi(\imath\omega)>0$ since $\gamma_1,\gamma_2>0$.

This formalization of the RRS problem makes it possible to apply the theory of optimal universal controllers, discussed in the previous section.
To design OUC~\eqref{eq:9} with the coefficients~\eqref{eq:13}, one has to choose $\rho(s)$ to be a scalar Hurwitz polynomial with $\deg\rho\ge \deg r+\deg\Delta$, whereas $r$
is a $2\times 2$ matrix polynomial that satisfies~\eqref{eq:15}.
By fixing $\rho(\imath\omega_j)$ and splitting the real and imaginary parts in the interpolation condition~\eqref{eq:15}, one obtains a pair of real-valued matrix equations for the coefficients of $r(s)$. The only exception is $j=N=p+1$: since $\omega_N=0$, the equation~\eqref{eq:15} is real-valued. Hence we get $1+2p$ equations for the coefficients of the polynomial $r$.
To satisfy them, the polynomial $r(s)$ should have $1+2p$ real-valued coefficients, i.e. it suffices to choose $\deg r= 2p$ and $\deg\rho\ge \deg\Delta+2p$.

The just described algorithm to design an OUC for the roll stabilization problem can be summarized as follows:
\begin{enumerate}
\item choose a Hurwitz polynomial $\rho(s)$ with $\deg\rho(s)\ge 2p+\deg\Delta$;
\item compute the matrices $R_j$ from~\eqref{eq:tf-special} (here $N=1+p$, $\omega_1,\ldots,\omega_p$ are the wave frequencies from~\eqref{eq:wave11}
and $\omega_N=\omega_{1+p}=0$);
\item compute $W_{yd}^0(\imath\omega_j)$ (see~\ref{app:Model});
\item find the real coefficients of the matrix polynomial $r(s)=r_0+\ldots+r_{2p}s^{2p}$ from~\eqref{eq:15a};
\item the controller~\eqref{eq:9} with the coefficients~\eqref{eq:13} provides optimality of~\eqref{eq:J1} for any uncertain amplitudes and phases.
\end{enumerate}

For the detailed derivation of the OUC controller one may represent the
transfer functions~\eqref{eq:vessel_model2} as follows
\begin{equation}\label{eq:exactTFs}
\begin{gathered}
W_{\varphi r}(s)=\frac{s b_{\varphi r}(s)}{a(s)},\, W_{\psi r}(s)=\frac{b_{\psi r}(s)}{a(s)},\\
W_{\varphi f}(s)=\frac{ s b_{\varphi f}(s)}{a(s)},\, W_{\psi f}(s)=\frac{b_{\psi f}(s)}{a(s)},
\end{gathered}
\end{equation}
In order to stabilize the vessel's heading, the autopilot controller is chosen to be
\begin{equation}\label{eq:apTF}
W_{ap} (s) = \frac{b_{ap}(s)}{a_{ap}(s)}.
\end{equation}

A straightforward computation of $W_{yu}^0(s)$, $W_{yd}^0(s)$ (see~\ref{app:Model}) shows that
\begin{equation*}
\begin{gathered}
W_{yu}^0(s)=\frac{1}{\Delta(s)}
\left[
\begin{matrix}
s a_{ap}(s) b_{\varphi r}(s)  & b^0_{\varphi u_2}(s)\\
a_{ap}(s) b_{\psi r}(s) & a_{ap}(s) b_{\psi f}(s)
\end{matrix}
\right],\\
W_{yd}^0(s)=
\left[
\begin{matrix}
-\frac{s b_{\varphi r}(s) b_{ap}(s)}{\Delta(s)} & \frac{1}{{\Delta(s)}} & \frac{s b_{\varphi r}(s) b_{ap}(s)}{\Delta(s)} \\
-\frac{a(s) a_{ap}(s)}{\Delta(s)} & 0 & \frac{a(s) a_{ap}(s)}{\Delta(s)}
\end{matrix}
\right],\\
\Delta(s) = a(s)a_{ap}(s) - b_{\psi r}(s) b_{ap}(s), \\
\begin{aligned}
	b^0_{\varphi u_2}(s) =
	&a_{ap}(s)
	b_{\varphi f}(s) +\\
	+& b_{ap}(s)
	\frac{b_{\varphi r}(s) b_{\psi f}(s) - b_{\varphi f}(s) b_{\psi r}(s)}{a(s)}.
\end{aligned}
\end{gathered}
\end{equation*}
Obviously, $\deg\Delta (s)= \deg a(s) + \deg a_{ap}(s)$.

The application of this procedure to a specific vessel's model is illustrated in the next section.

\section{Numerical simulation}\label{sec:sim}

In this section we consider a numerical example to illustrate the proposed approach. 

\subsection{Vessel's motion and wave disturbance}

{To design the linear controllers, we consider the vessel's 4-DoF (surge, sway, roll and yaw) model  from~\citep[Appendix B]{perez2006ship}. This maneuvering model of a multipurpose naval vessel is implemented in the Marine Systems Simulator (MSS) Toolbox~\citep{mss_toolbox} in our numerical simulations.
}
% For simulations, we consider the vessel's 4-DoF model  from~\citep[Appendix B]{perez2006ship} linearized at the constant speed 8 m/s. 
The vessel has one rudder and two synchronous fins.
Linearizing the model at the speed $8$m/s, the coefficients of the transfer functions~\eqref{eq:exactTFs}  are as follows~\citep[Appendix B]{perez2006ship}
\[
\begin{gathered}
a(s)= s (s+0.4375)(s+0.04404)(s^2+0.2164s+1.31),\\
b_{\varphi r}(s)= -0.159(s-0.4919)(s+0.3005),\\
b_{\psi r}(s)= -0.078(s+0.1785)(s^2 + 0.2586s + 1.324),\\
b_{\varphi f}(s)= 0.402(s+0.4501)(s+0.03056),\\
b_{\psi f}(s)= -0.006(s-0.9642)(s^2+0.1974s+0.2361),\\
\end{gathered}
\]
For this simulation we assume that the stabilizing autopilot~\eqref{eq:apTF} has the following form
\[
a_{ap}(s)= (s+10),\quad b_{ap}(s)=57(s+0.5263).
\]
{Whereas the controller design is based on this linearized model, our simulations take into account the nonlinear dynamics of the rudder's and fins' steering machines~\citep[5.6]{perez2006ship} with the maximal angles $40$ degrees (rudder) 
and $35$ degrees (fins) and the maximal rates {$5$} degree/s and $25$ degree/s respectively.~\citep[Appendix B.4]{perez2006ship}.}

\begin{figure*}[h]
\centering
\begin{subfigure}{.49\textwidth}
	\centering
	\includegraphics[width=0.95\linewidth]{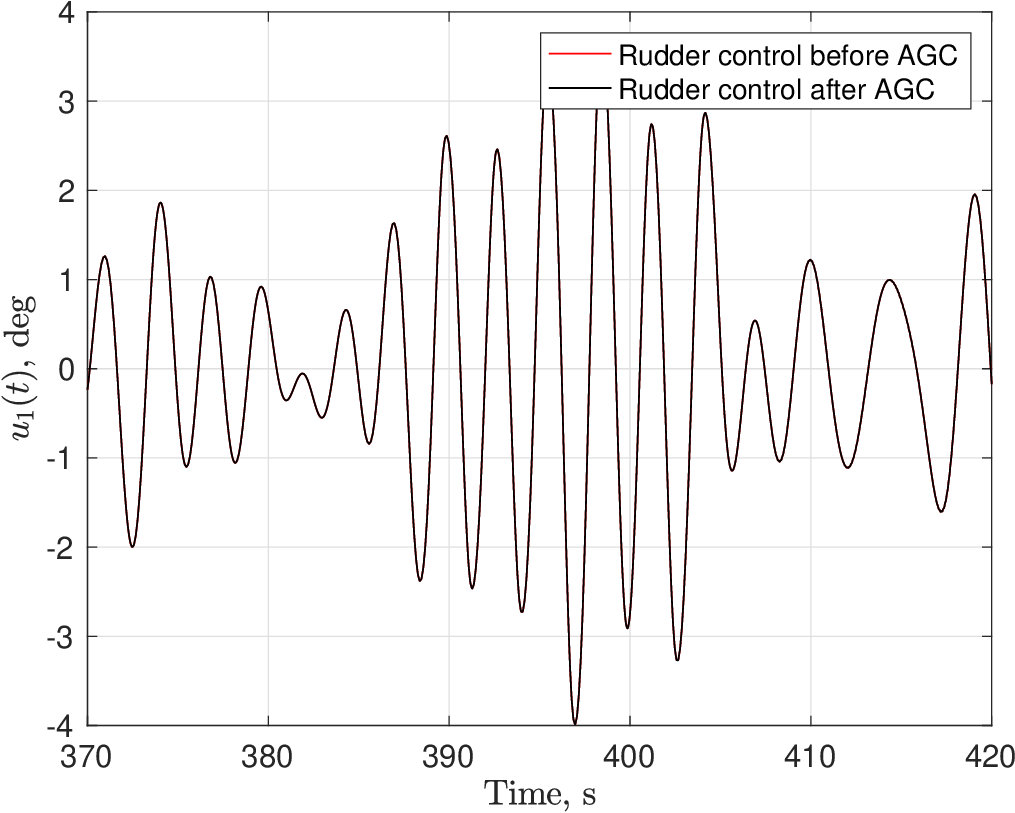}
	%\subcaption{Rudder command before and after AGC}
\end{subfigure}%\\[3mm]
\begin{subfigure}{.49\textwidth}
	\centering
	\includegraphics[width=0.95\linewidth]{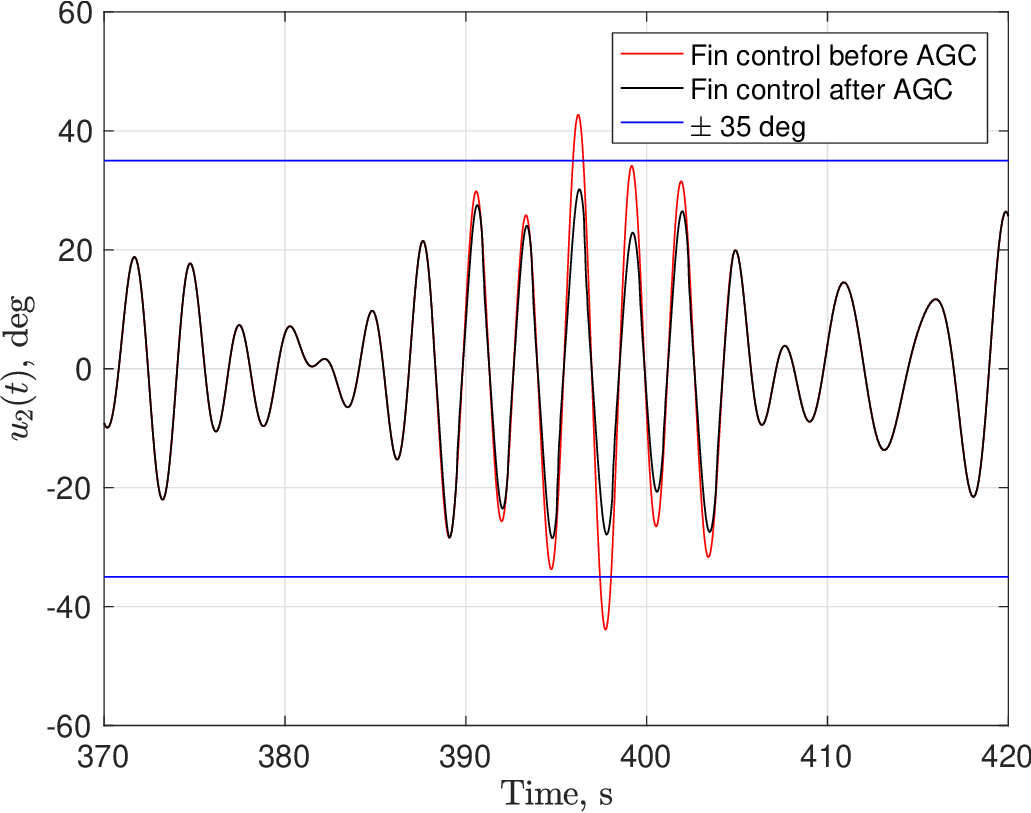}
	%\subcaption{Fin angle command before and after AGC}
\end{subfigure}
\caption{Commands for the rudder (not saturated in the experiment) and fin angle (saturation prevented by AGC)}\label{fig:agc}
\end{figure*}

{To cope with saturations, we use the heuristical approach, called the automatic gain control (AGC)~\citep{van1987rudder,Lauvdal1998}
that decreases the actuator command in a smart way to ensure that saturation (of the actuator angle or rate) never occurs. In Fig.~\ref{fig:agc} we compare the commands from our OUC (whose design will be specified in Subsection~5.3) 
before and after the AGC algorithm. The rudder is not saturated, in this case the command remains unchanged. The fin angle's saturation is prevented by AGC.}

\begin{figure*}
\centering
\begin{subfigure}{0.24\textwidth}
	\centering
	\includegraphics[width=0.95\linewidth]{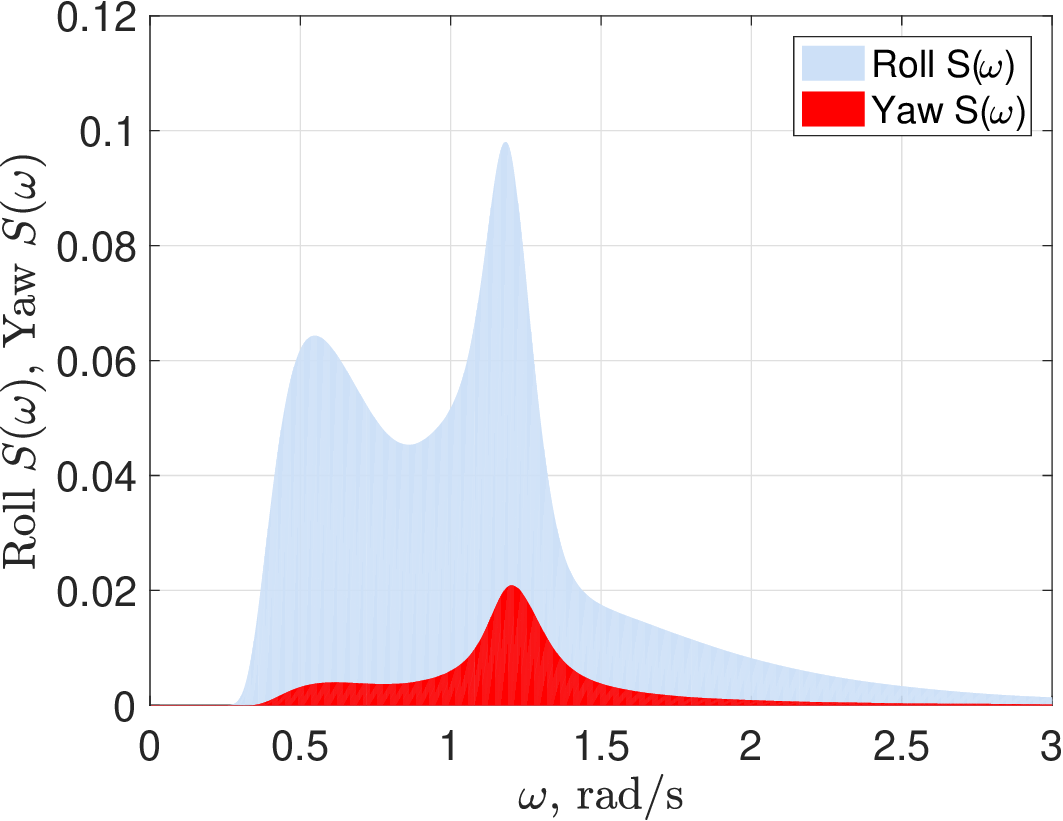}
	\subcaption{Cases 1-3}
\end{subfigure}
\begin{subfigure}{0.24\textwidth}
	\centering
	\includegraphics[width=0.95\linewidth]{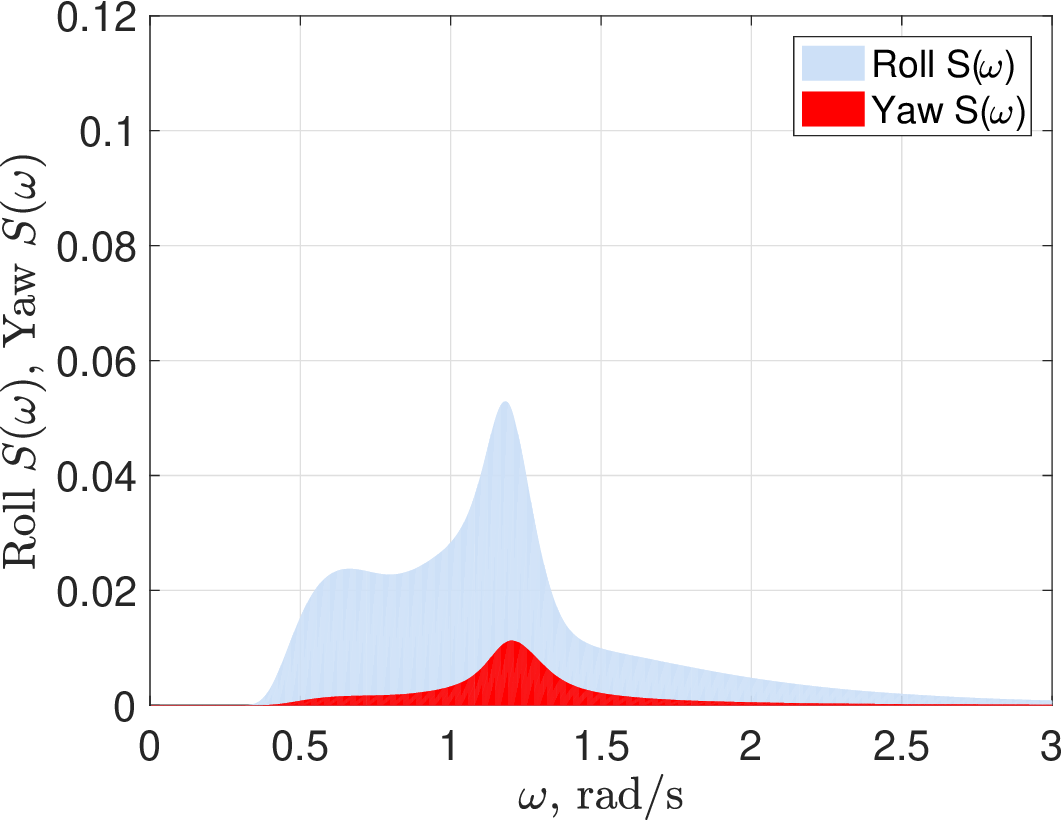}
	\subcaption{Case 4}
\end{subfigure}
\begin{subfigure}{0.24\textwidth}
	\centering
	\includegraphics[width=0.95\linewidth]{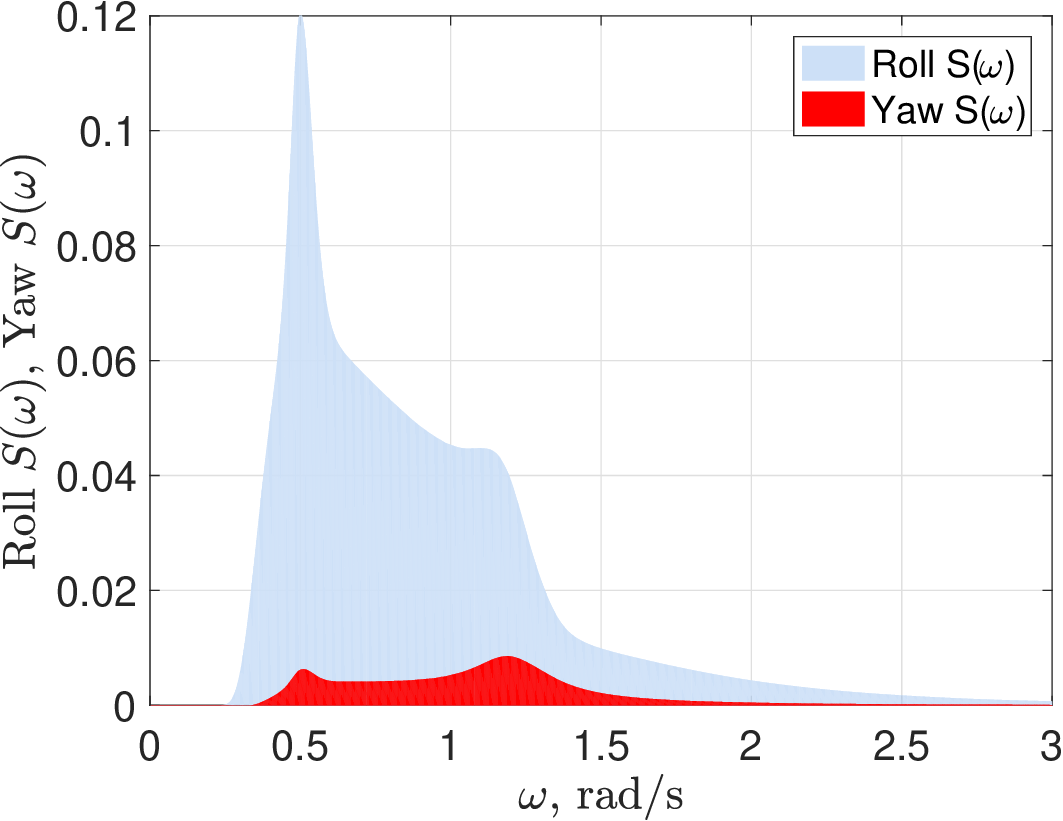}
	\subcaption{Case 5}
\end{subfigure}%\\[5mm]
\begin{subfigure}{0.24\textwidth}
	\centering
	\includegraphics[width=0.95\linewidth]{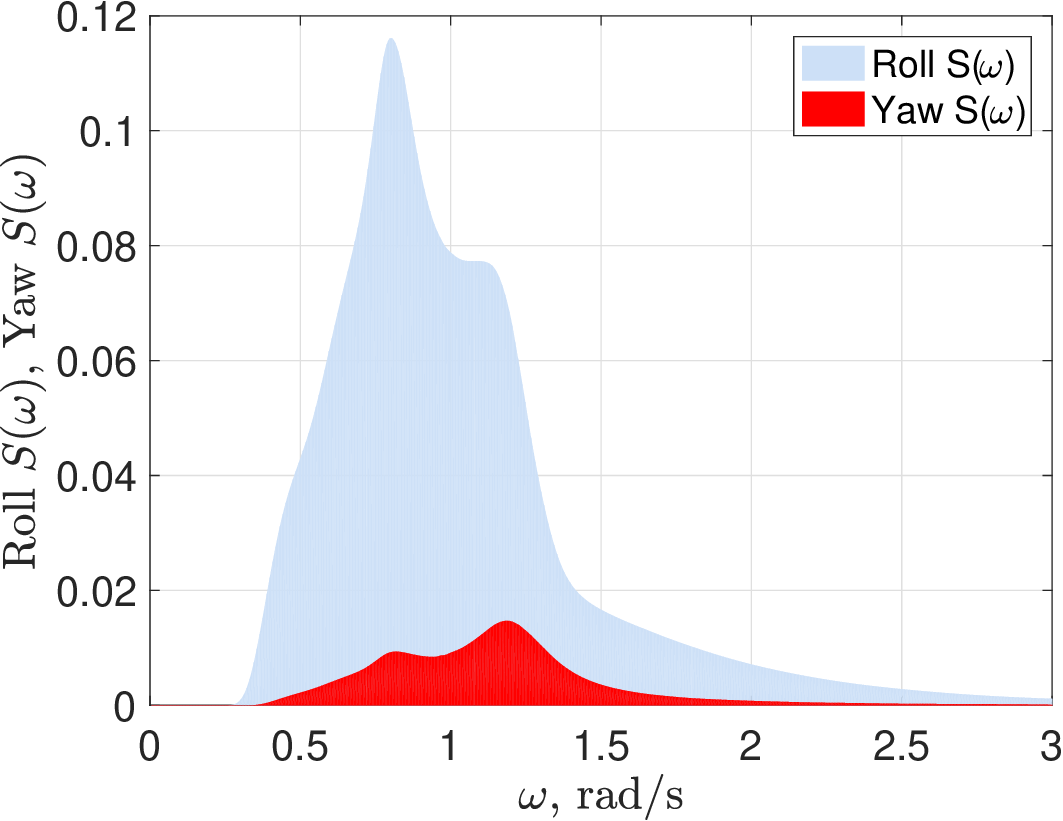}
	\subcaption{Case 6}
\end{subfigure}
\caption{Power spectral densities of the wave disturbance in Cases 1-6.}\label{fig:wavepsd}
\end{figure*}

To obtain the proper time series of the polyharmonic approximation of the irregular wave~\eqref{eq:wave11} we use the methodology presented in~\citep[Sect. 4.2.5]{Perez2012129}. The resulting realization is obtained for the \emph{long-crested} irregular sea in beam seas. The response amplitude operator has been taken from~\citep[Table B.9.]{Perez2012129}, where the number of sinusoidal components is  $1000$. {We consider the JONSWAP spectrum~\citep[Section~4.2.1]{fossen1994guidance}, which is characterized by two parameters: the significant wave height and the peak value of the spectrum (peak frequency). We consider two different significant wave heights ($1.5$m and $3$m) and three different peak values ($1.15$ rad/s, $0.8$ rad/s, $0.5$ rad/s). 
Also, we compare the controllers' behavior at three different speeds: $8$m/s (the linearization point of the model at which the controller is designed), $5$m/s (medium speed) and $1$m/s (low speed at which the rudders and fins are limited with small inflow velocity). In total, we consider 6 different scenarios (Table~\ref{tab:1}).
The power spectra are shown in Fig.~\ref{fig:wavepsd}a-d).}

%In order to design the OUC following the procedure from the previous section and to demonstrate its efficiency, 
We use ``rough'' approximations of the signal~\eqref{eq:wave11} taking only one ``dominating'' sinusoidal components. {For the frequencies of the sinusoidal signal, we choose $\omega_1=1.15$ rad/s and
$\omega_2=0.5$ rad/s. To evaluate the robustness of OUC against unspecified harmonics, in Case 6 we consider the peak frequency $0.8$rad/s which is different from $\omega_1$ and $\omega_2$.}

%\begin{minipage}{\textwidth}
%\end{minipage}

{
\begin{center}
\begin{table}
\caption{Parameters of simulations \label{tab:1}}
\begin{tabularx}{0.45\textwidth}{|X|X|X|X|}
% { | >{\centering\arraybackslash}X | >{\centering\arraybackslash}X | >{\centering\arraybackslash}X | >{\centering\arraybackslash}X | }
% \caption{Parameters of simulations}\label{tab:1}\\
% \caption{Table caption \label{tab:table_label}}
\hline
 Case & Vessel speed, m/s & Wave Height, m & Peak value, rad/s\\
%  \endfirsthead
 \hline\hline
 1 & 8 & 3 & 1.15 \\ 
 \hline
 2 & 5 & 3 & 1.15 \\
 \hline
 3 & 1  & 3  & 1.15 \\
 \hline
 4 & 8  & 1.5 & 1.15 \\
 \hline
 5 & 8  & 3 & 0.5 \\
 \hline
 6 & 8  & 3 & 0.8 \\
 \hline
\end{tabularx}
\end{table}
\end{center}
}

\subsection{The standard controllers}

We compare the OUC with two other types of linear controllers. The first of these controllers is the classical  \emph{LQR} controller~\citep[Appendix D]{fossen1994guidance}, 
{designed to optimize the cost function}
\begin{equation}\label{eq.cost-lqr}
J_{LQR} = \int\limits_{0}^{\infty}\left[
\alpha' e_\varphi^2 + \beta' e_\psi^2 + \gamma_1' u_1^2 + \gamma_2' u_2^2
\right]dt.    
\end{equation}
{in the \emph{absence} of disturbances. We choose the parameters $\alpha'=5$, $\beta'=1$, $\gamma_1'=0.01$, $\gamma_2'=0.001$.}

The conventional loop shaping controller~\citep{VANAMERONGEN1990679} has been chosen as another algorithm for comparison. This method ignores the yaw dynamics, working directly with the transfer function $W^0_{\vp {u_2}}$ (see~\ref{app:Model}). For the known dominating frequency of the disturbance, we select the structure of the notch filter centered at that point to damp the frequencies around it. The controller takes the following form
\[
W_{c}(s) = \frac{u_1(s)}{e_\vp(s)} = -10\frac{s^2 + 0.2(1.15)s+1.15^2}{(s+1.15)^2}.
\]

\subsection{The OUC design}

The coefficients of the cost functional~\eqref{eq:J1} are chosen as $\alpha = 5$, $\beta = 1$, $\gamma_1 = 10$ and $\gamma_2 = 2$. It should be noted that, in spite of similar cost functions, the behaviors of the two controllers are very different. Actually, the cost function~\eqref{eq.cost-lqr} is in principle finite only for the disturbances vanishing as $t\to\infty$, and its optimality for $d\equiv 0$ does not guarantee any optimal performance for the polyharmonic disturbance. For this reason, the choice of the coefficients $\alpha,\beta,\gamma_1,\gamma_2$ is a delicate issue. We have tuned them in such a way that the LQR and OUC provide (approximately) same quality of course keeping (see the heading errors in Table~\ref{tab:stdYaw}).

{In the design procedure from the previous section, we should choose the Hurwitz polynomial of the order
{$\deg\rho>\deg\Delta+2p=10$. The polynomial $\rho(s)$ is chosen as follows
\begin{equation}\label{eq.rho}
\rho(s)=\Delta_{LQR}(s)(s+\eta)(s^2+2\xi\omega_1s+\omega_1^2)(s^2+2\xi\omega_2s+\omega_2^2).
\end{equation}
Here $\omega_0=0$, $\omega_1=0.5$, $\omega_2=1.15$ are the frequencies of the wave disturbance, $\xi=1/\sqrt{2}$, $\eta=2$ and
\[
\begin{aligned}
\Delta_{LQR}(s)=s^6+17.2s^5+101.3s^4+300.1s^3+\\+201.3s^2+52s+5.
\end{aligned}
\]
is the characteristic polynomial of the closed-loop system, which corresponds to the LQR controller described above. The reasons to choose this polynomial are discussed below in Subsection~\ref{subsec:choice}. 
}

\subsection{Simulation results}

\begin{figure*}[h]
\centering
\begin{subfigure}{0.49\textwidth}
\includegraphics[width=0.95\linewidth]{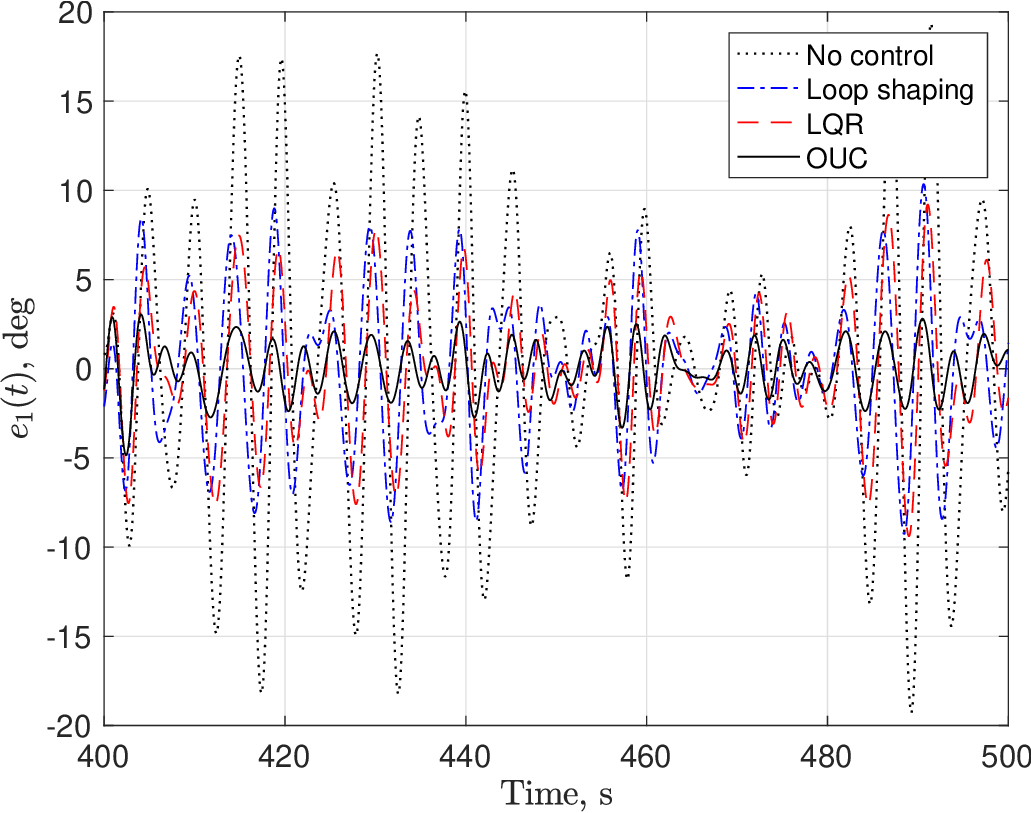}
\end{subfigure}
\begin{subfigure}{0.49\textwidth}
\includegraphics[width=0.95\linewidth]{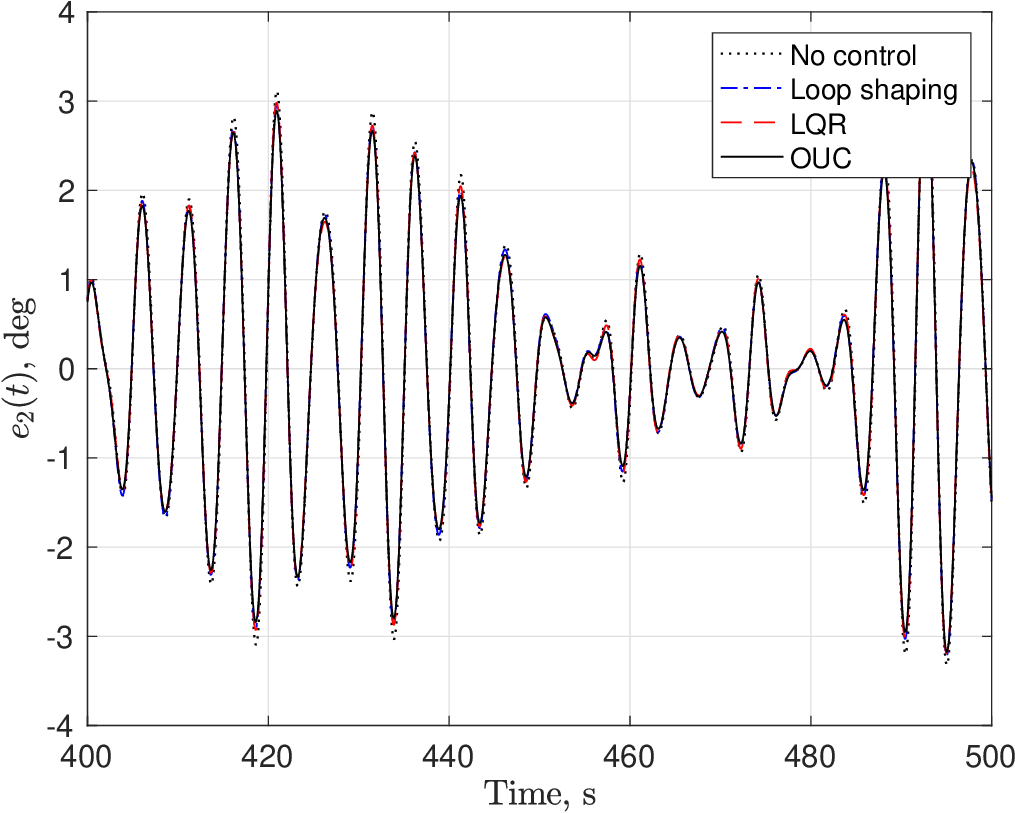} 
\end{subfigure}\\[5mm]
\begin{subfigure}{0.49\textwidth}
\includegraphics[width=0.95\linewidth]{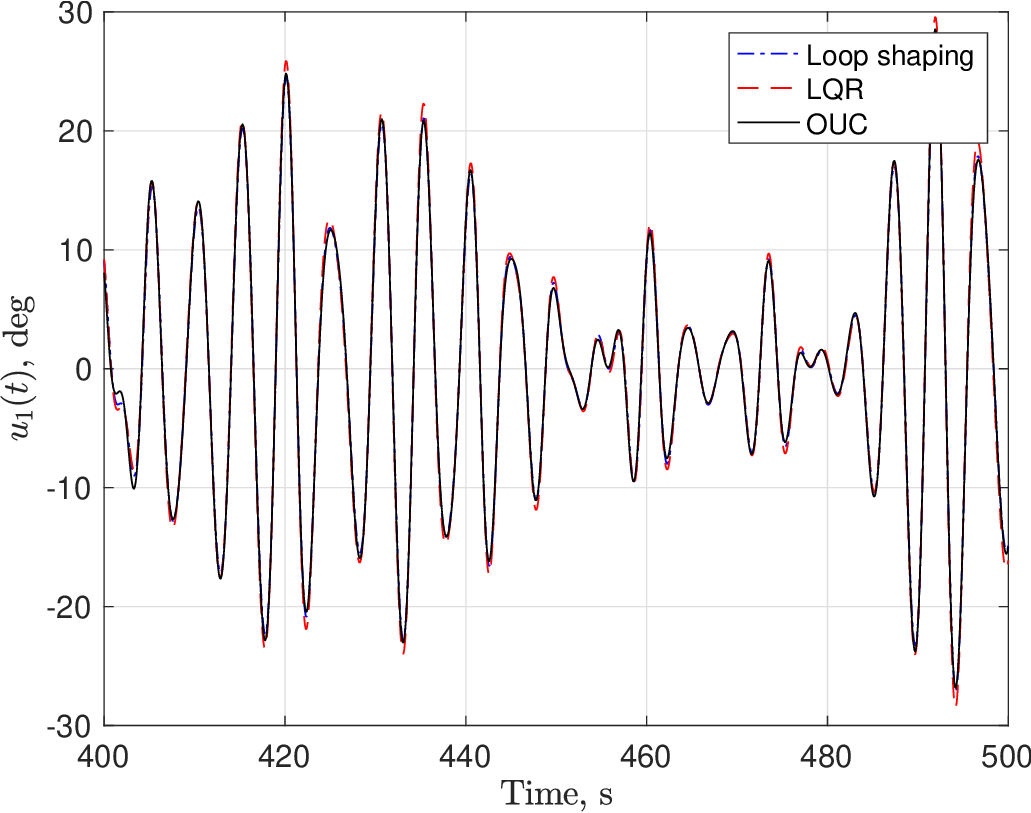} 
\end{subfigure}
\begin{subfigure}{0.49\textwidth}    
\includegraphics[width=0.95\linewidth]{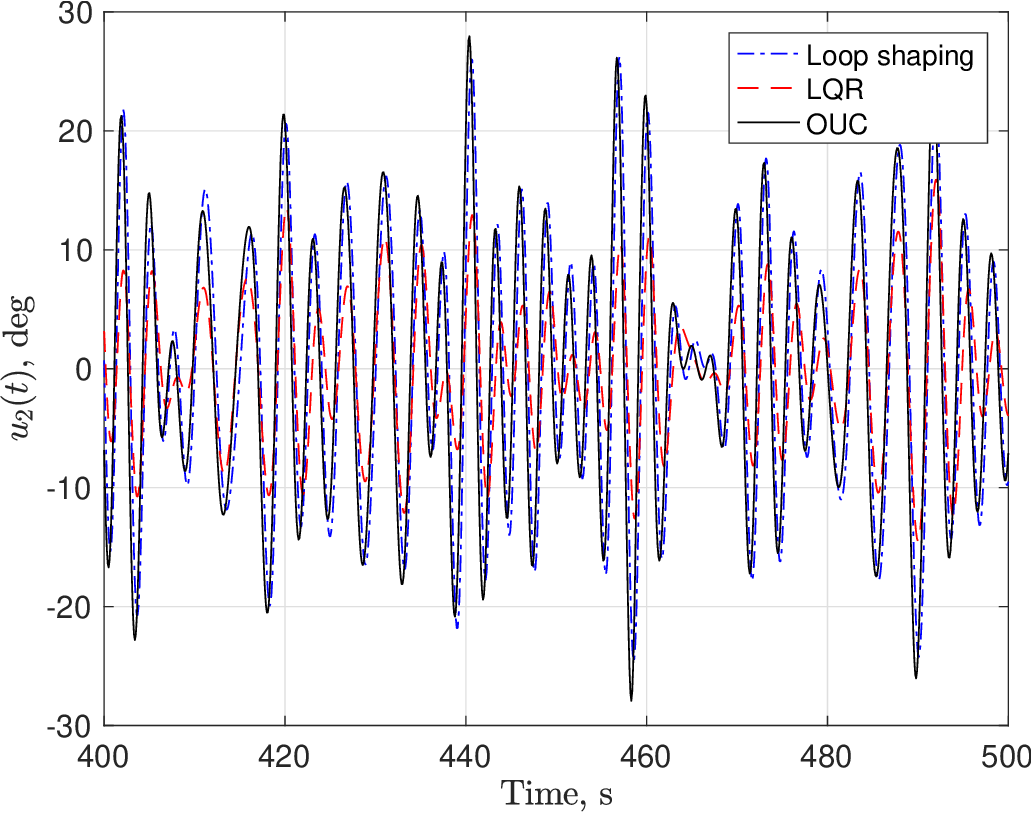}
\end{subfigure}
\caption{The result of simulation in Case 1: roll angle $e_1$, heading error $e_2$, rudder and fin angles ($u_1,u_2$).}\label{fig:case1}
\end{figure*}
\begin{figure*}[h]
\centering
\begin{subfigure}{0.49\textwidth}
\includegraphics[width=0.95\linewidth]{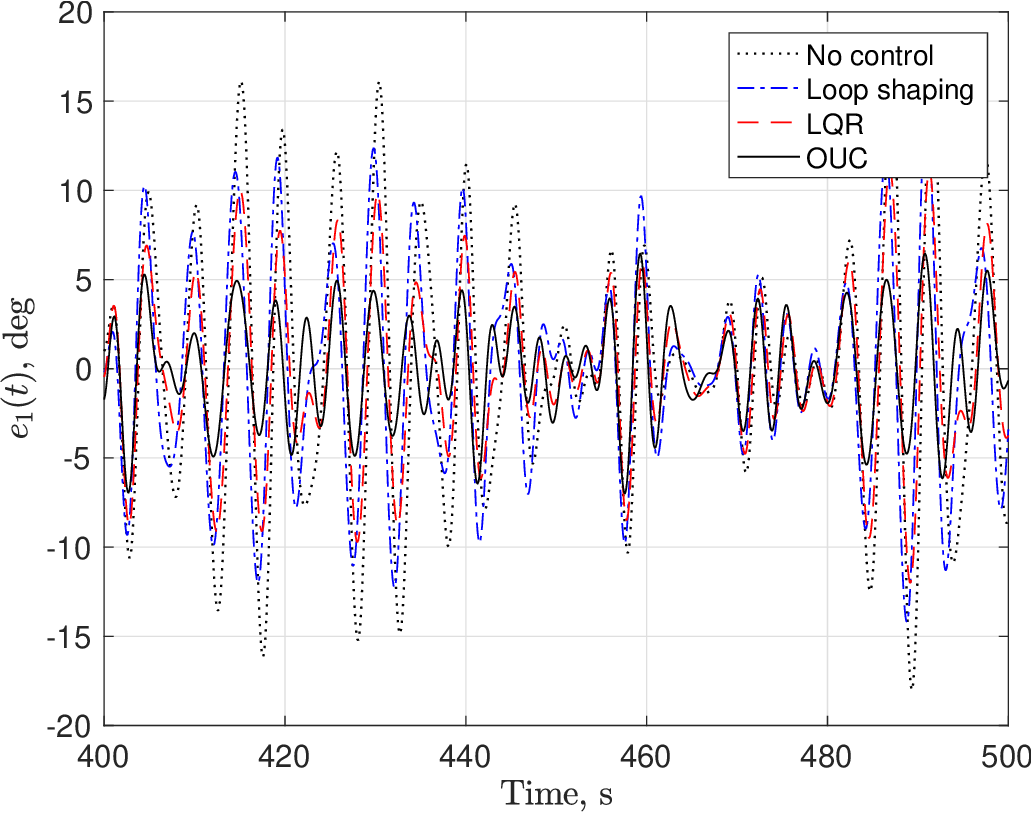}
\end{subfigure}
\begin{subfigure}{0.49\textwidth}
\includegraphics[width=0.95\linewidth]{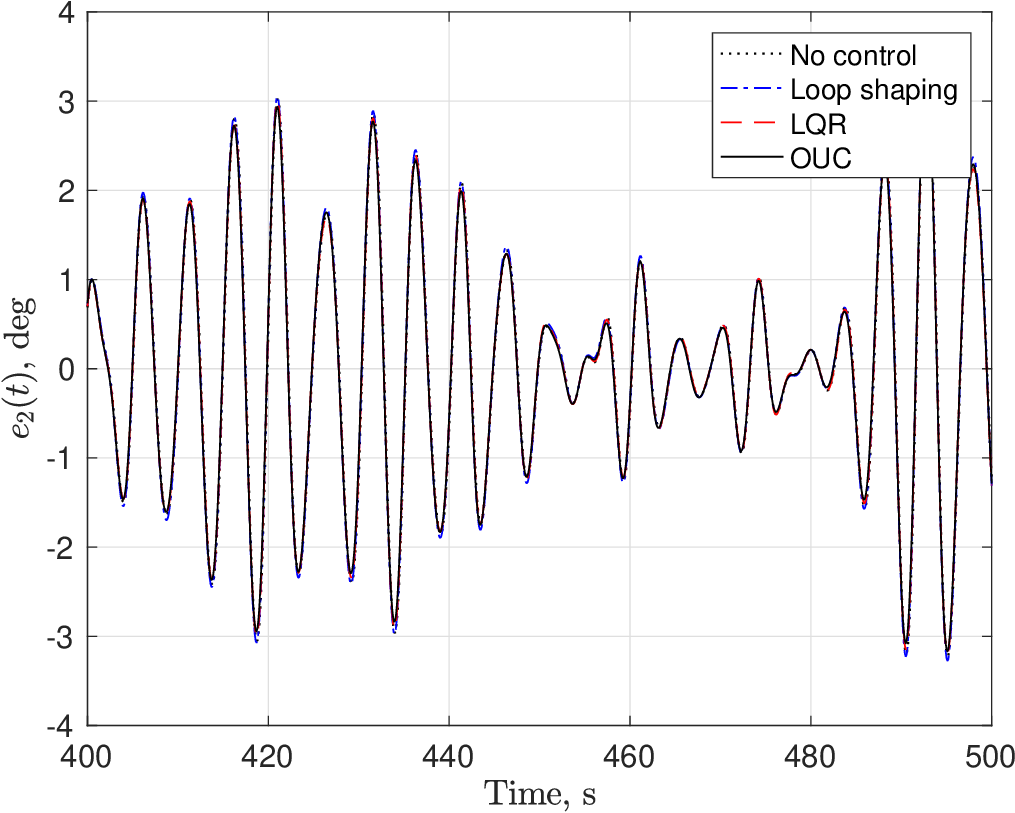} 
\end{subfigure}\\[5mm]
\begin{subfigure}{0.49\textwidth}
\includegraphics[width=0.95\linewidth]{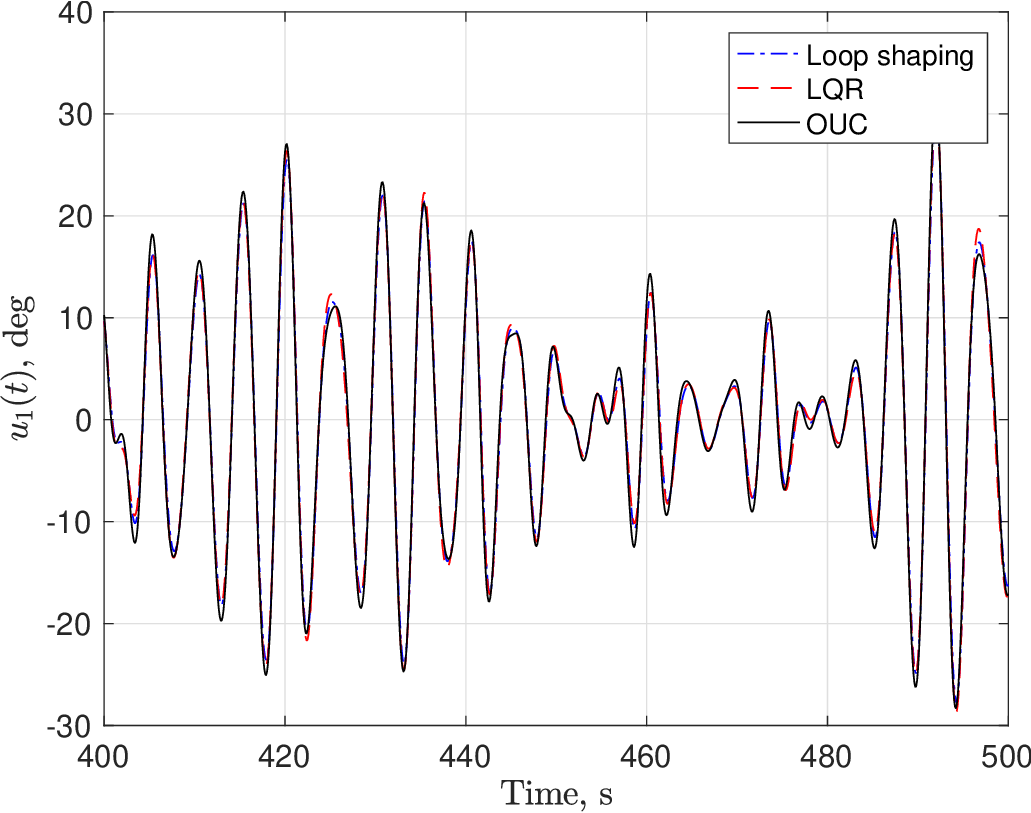} 
\end{subfigure}
\begin{subfigure}{0.49\textwidth}
\includegraphics[width=0.95\linewidth]{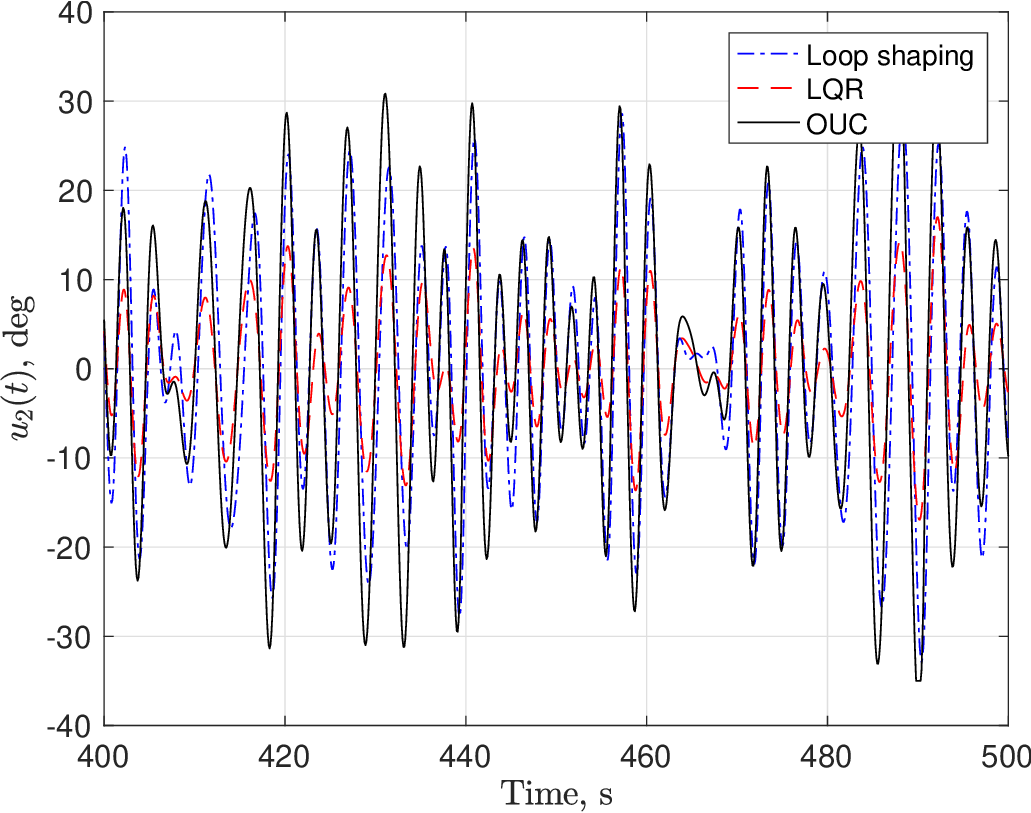}
\end{subfigure}
\caption{The result of simulation in Case 2: roll angle $e_1$, heading error $e_2$, rudder and fin angles ($u_1,u_2$).}\label{fig:case2}
\end{figure*}
\begin{figure*}[h]
\centering
\begin{subfigure}{0.49\textwidth}
\includegraphics[width=0.95\linewidth]{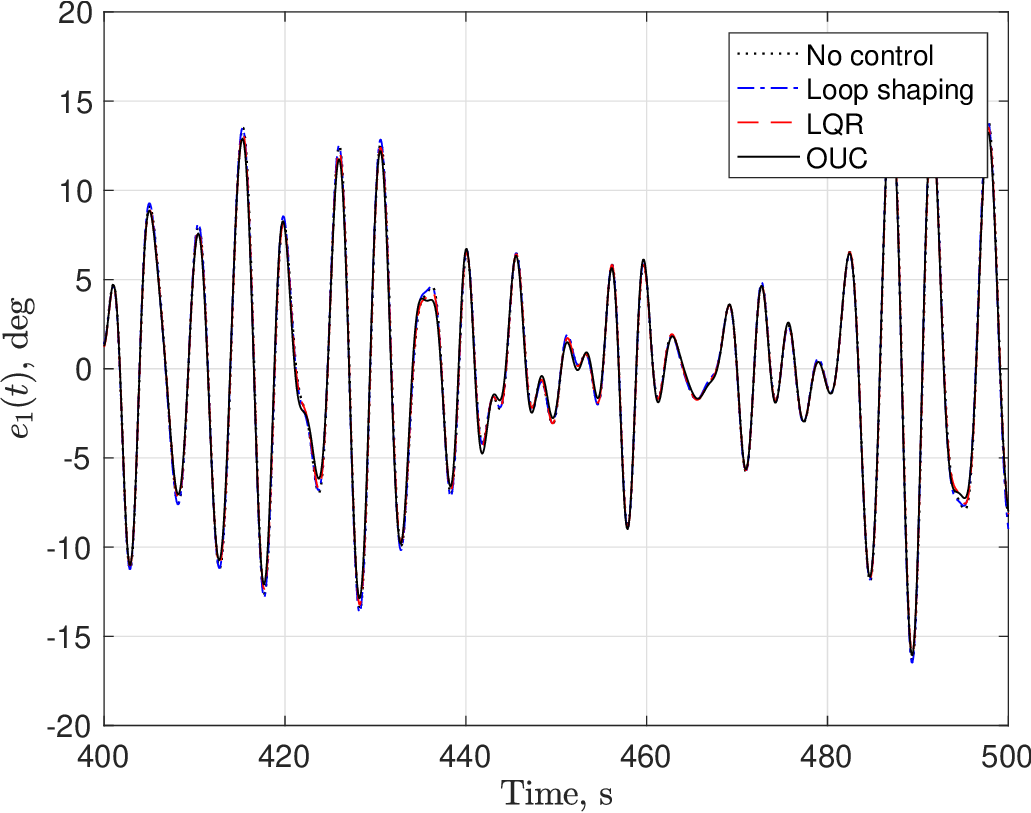}
\end{subfigure}
\begin{subfigure}{0.49\textwidth}
\includegraphics[width=0.95\linewidth]{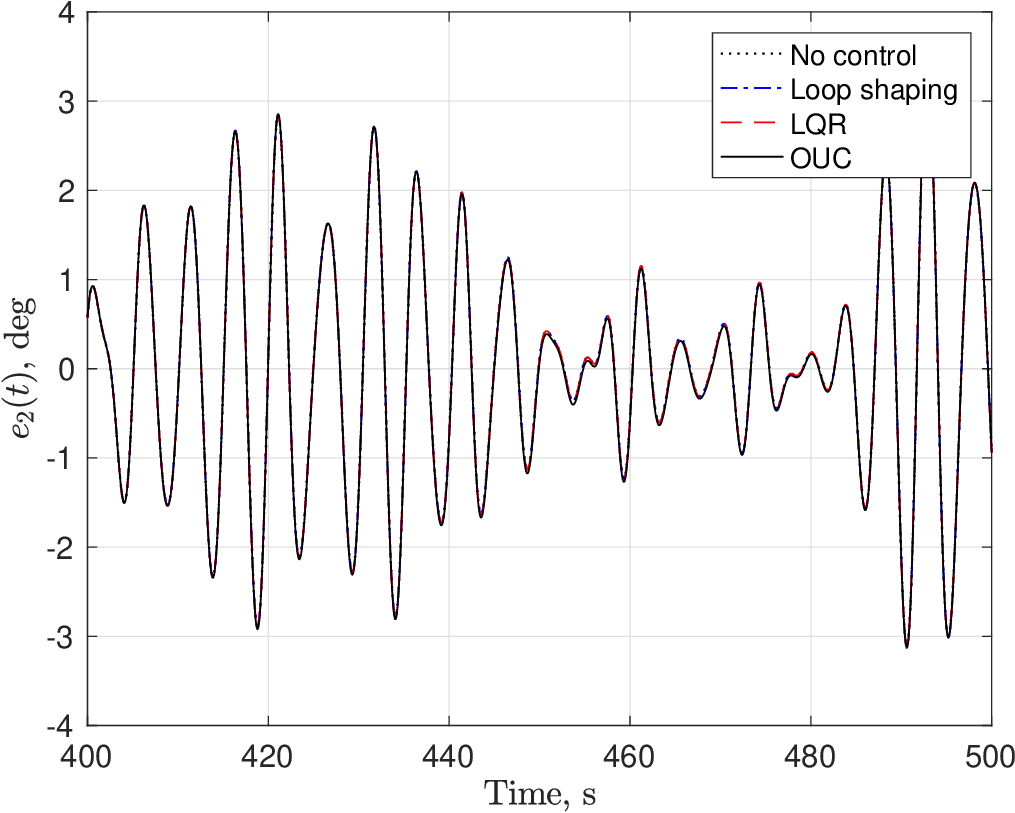} 
\end{subfigure}\\[5mm]
\begin{subfigure}{0.49\textwidth}
\includegraphics[width=0.95\linewidth]{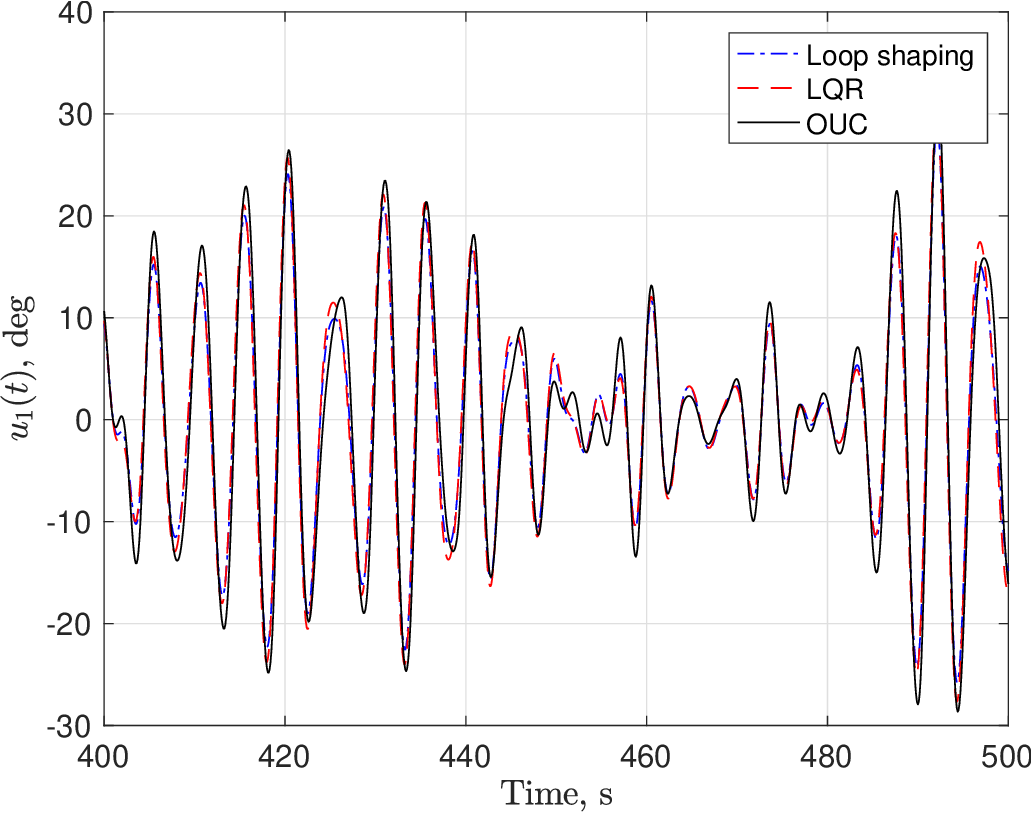} 
\end{subfigure}
\begin{subfigure}{0.49\textwidth}
\includegraphics[width=0.95\linewidth]{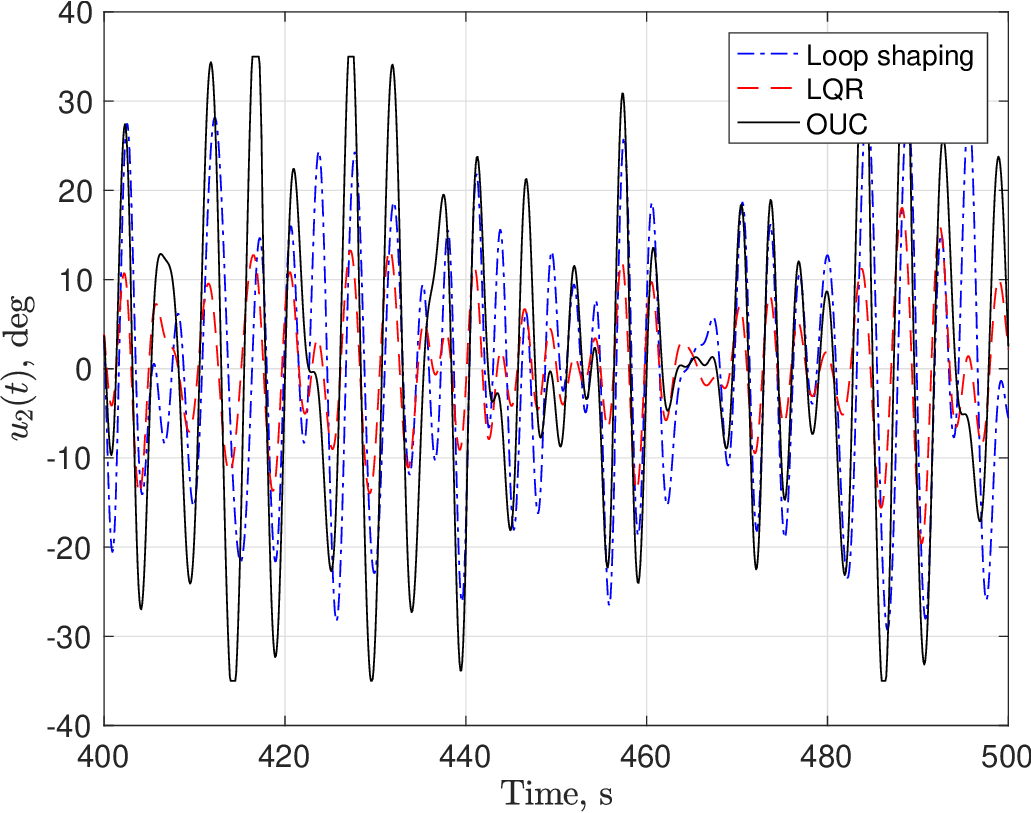}
\end{subfigure}
\caption{The result of simulation in Case 3: roll angle $e_1$, heading error $e_2$, rudder and fin angles ($u_1,u_2$).}\label{fig:case3}
\end{figure*}
\begin{figure*}[h]
\centering
\begin{subfigure}{0.49\textwidth}
\includegraphics[width=0.95\linewidth]{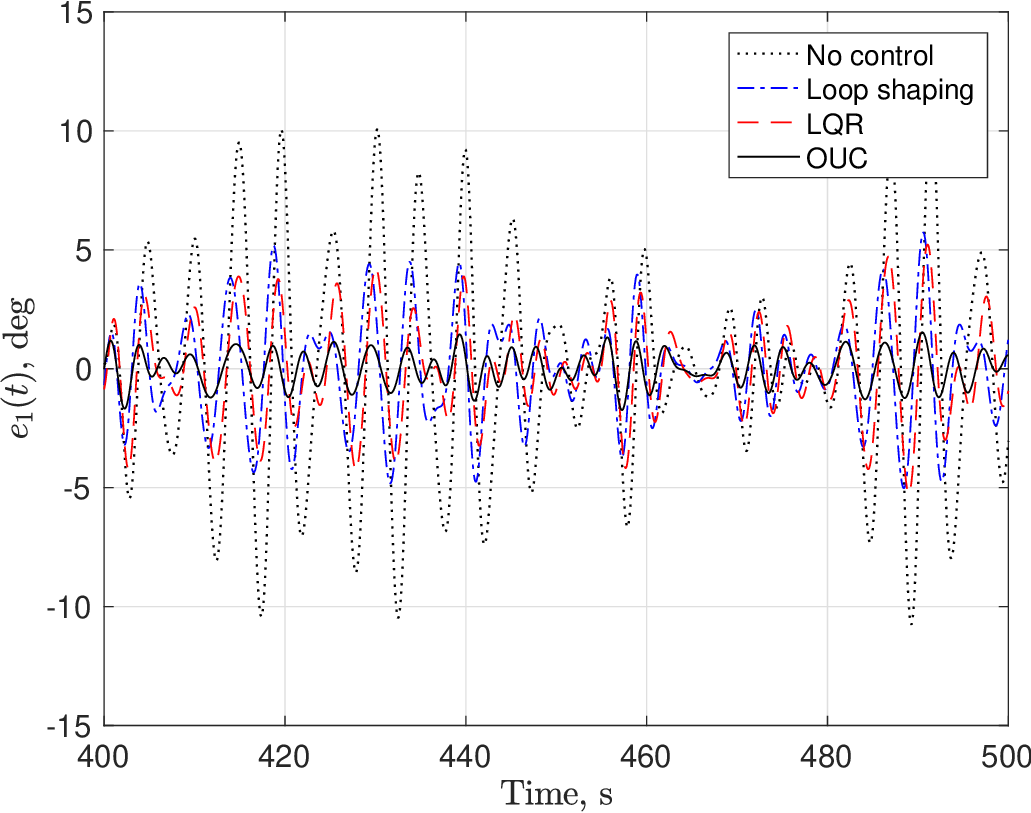}
\end{subfigure}
\begin{subfigure}{0.49\textwidth}
\includegraphics[width=0.95\linewidth]{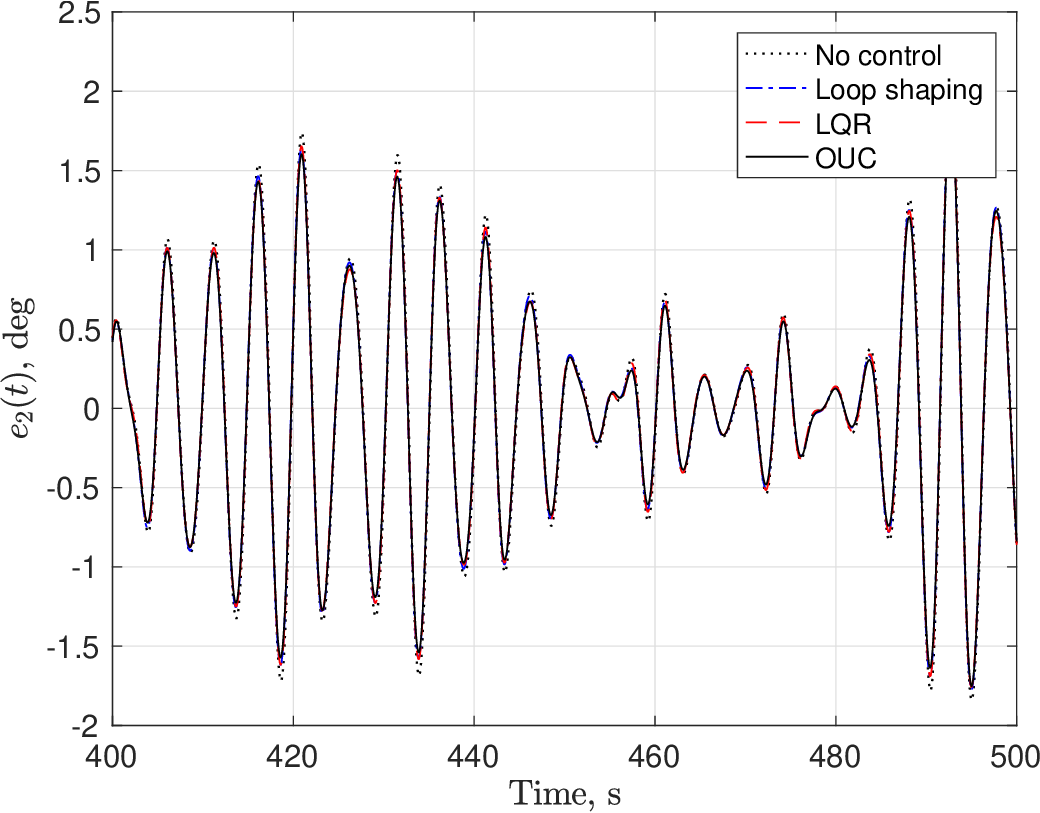} 
\end{subfigure}\\[5mm]
\begin{subfigure}{0.49\textwidth}
\includegraphics[width=0.95\linewidth]{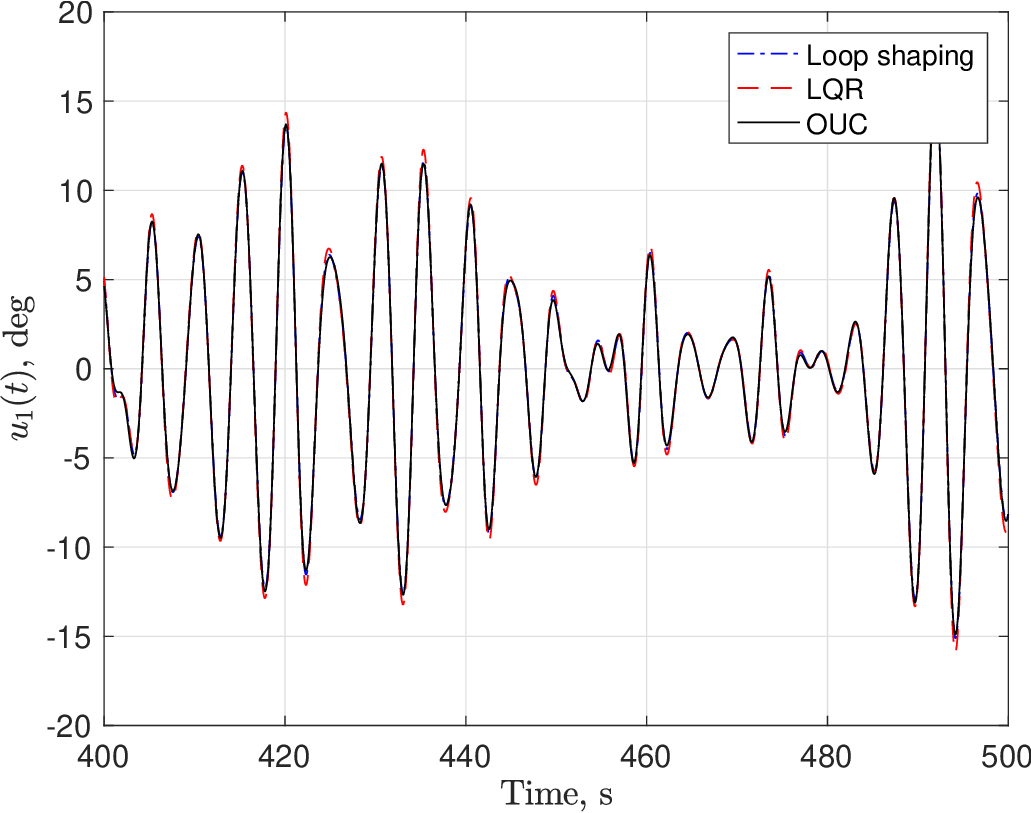} 
\end{subfigure}
\begin{subfigure}{0.49\textwidth}
\includegraphics[width=0.95\linewidth]{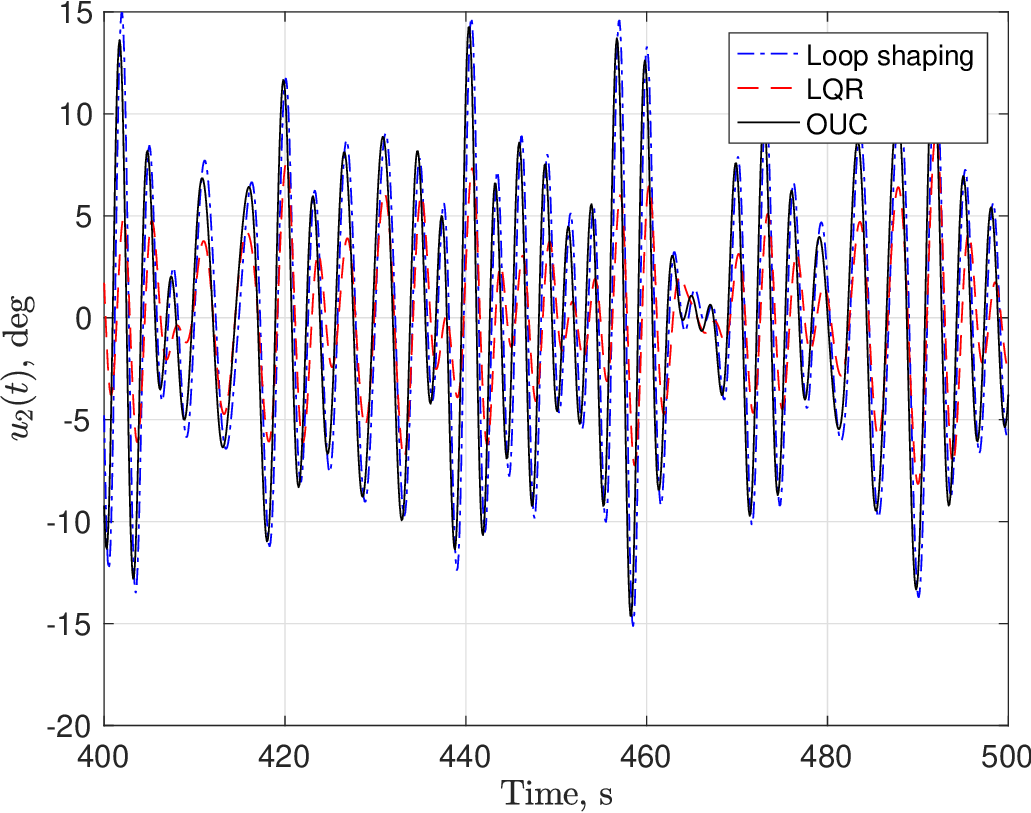}
\end{subfigure}
\caption{The result of simulation in Case 4: roll angle $e_1$, heading error $e_2$, rudder and fin angles ($u_1,u_2$).}\label{fig:case4}
\end{figure*}
\begin{figure*}[h]
\centering
\begin{subfigure}{0.49\textwidth}
\includegraphics[width=0.95\linewidth]{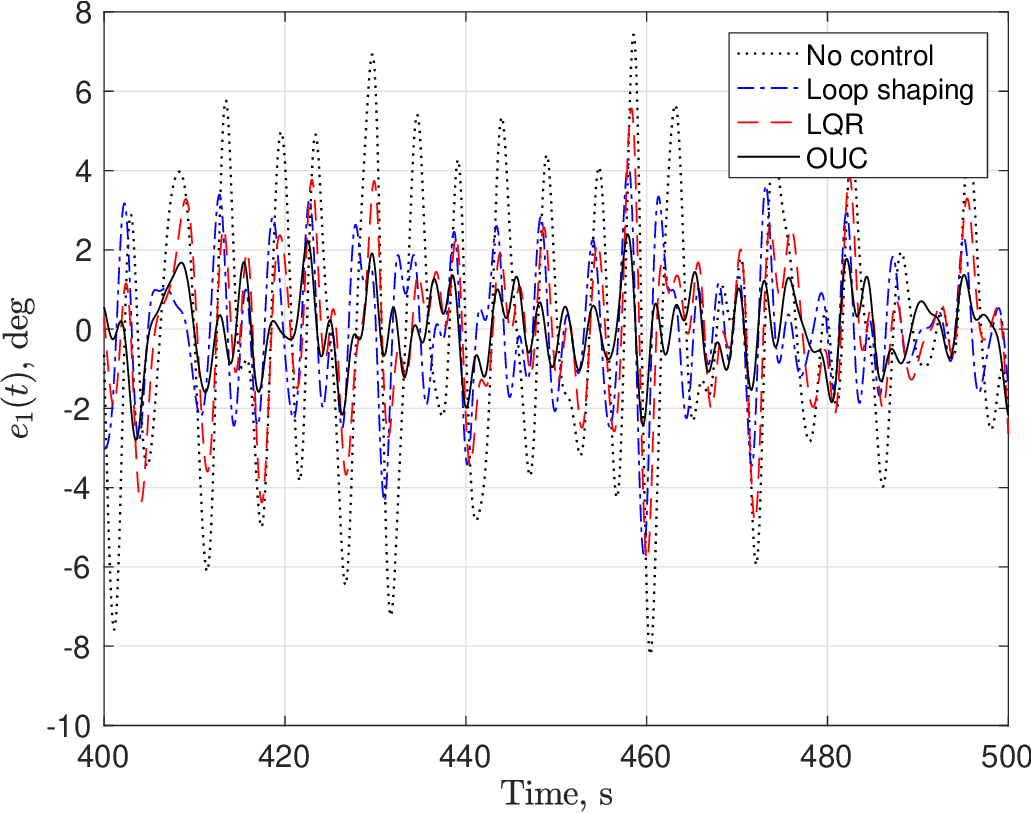}
\end{subfigure}
\begin{subfigure}{0.49\textwidth}
\includegraphics[width=0.95\linewidth]{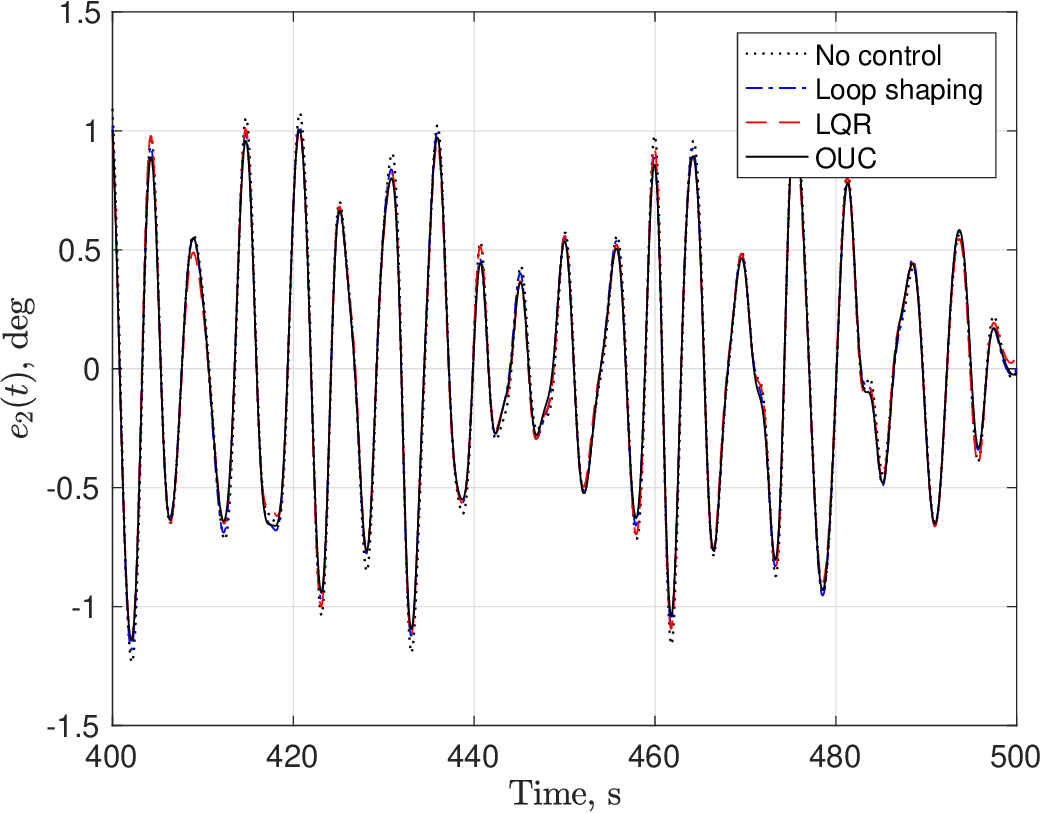} 
\end{subfigure}\\[5mm]
\begin{subfigure}{0.49\textwidth}
\includegraphics[width=0.95\linewidth]{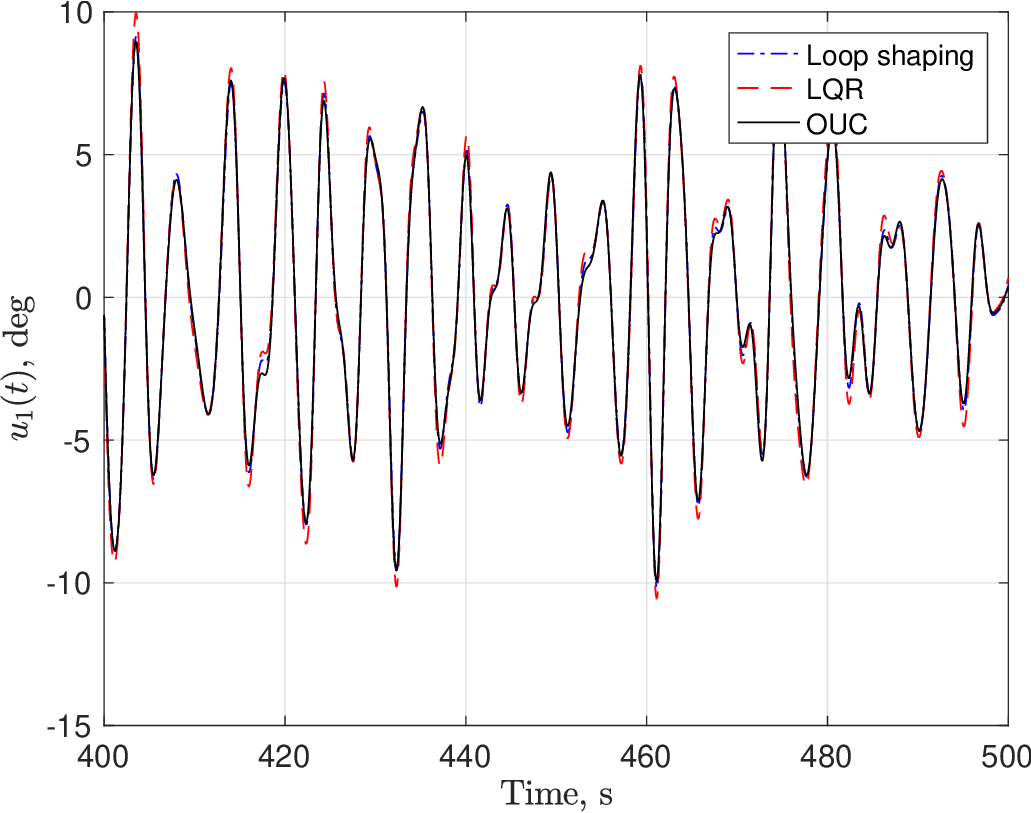} 
\end{subfigure}
\begin{subfigure}{0.49\textwidth}
\includegraphics[width=0.95\linewidth]{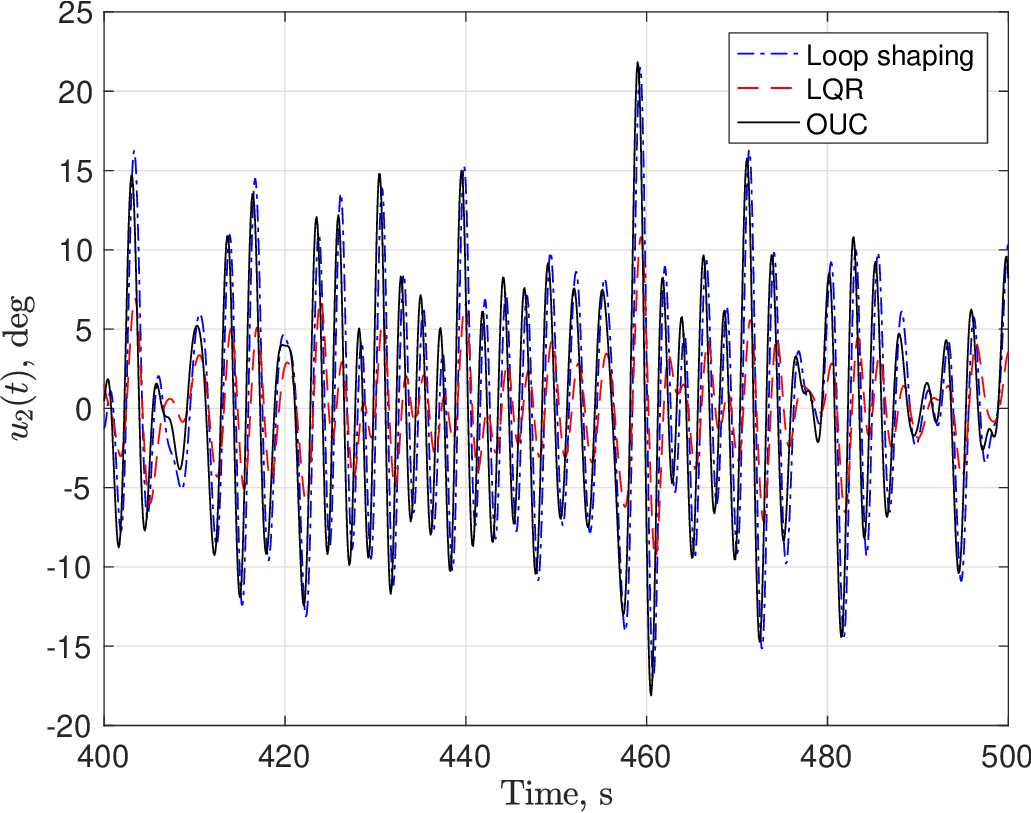}
\end{subfigure}
\caption{The result of simulation in Case 5: roll angle $e_1$, heading error $e_2$, rudder and fin angles ($u_1,u_2$).}\label{fig:case5}
\end{figure*}
\begin{figure*}[h]
\centering
\begin{subfigure}{0.49\textwidth}
\includegraphics[width=0.95\linewidth]{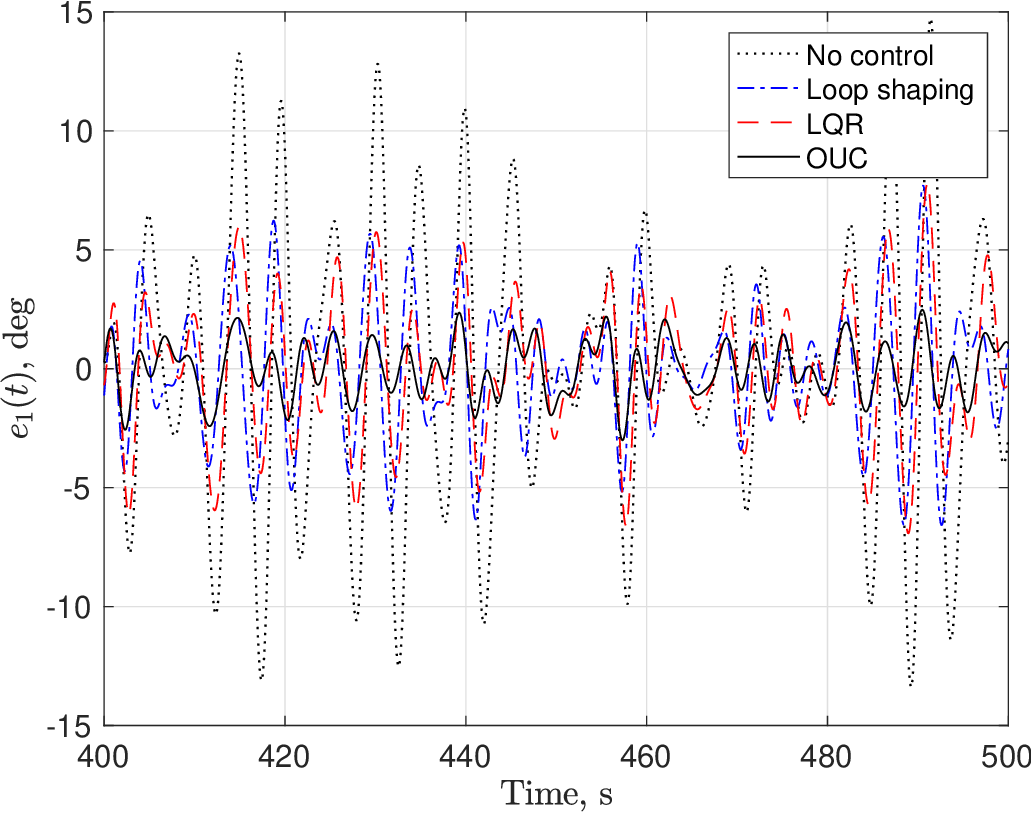}
\end{subfigure}
\begin{subfigure}{0.49\textwidth}
\includegraphics[width=0.95\linewidth]{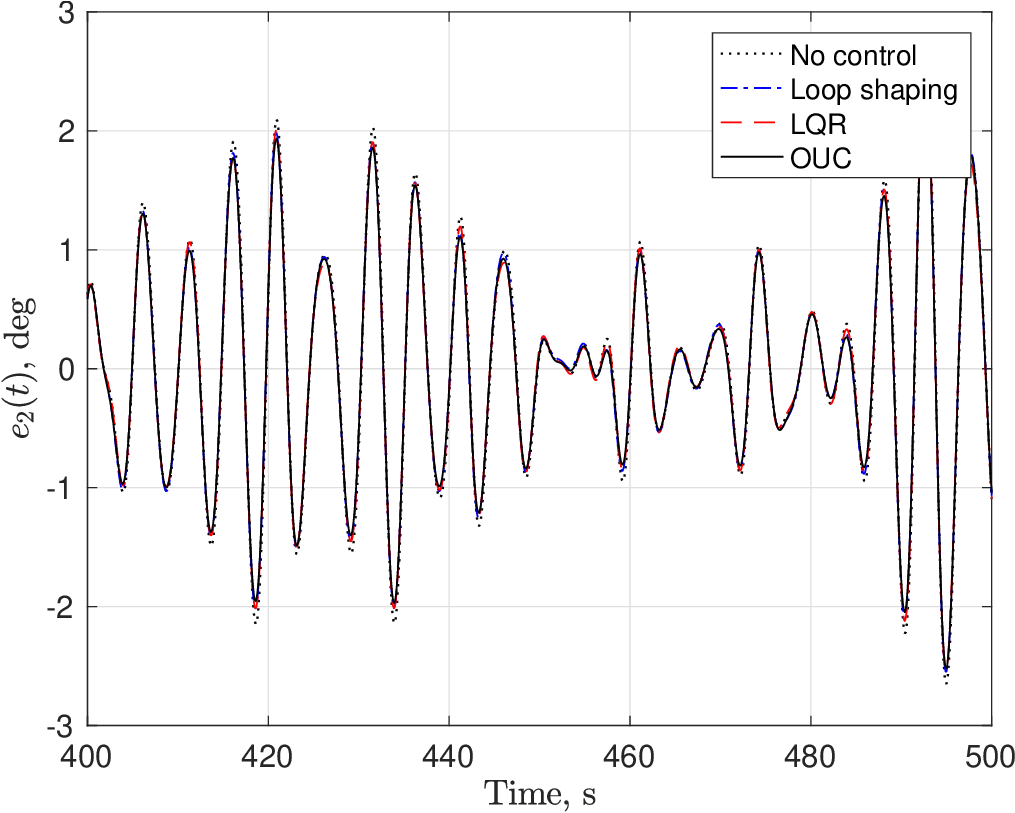} 
\end{subfigure}\\[5mm]
\begin{subfigure}{0.49\textwidth}
\includegraphics[width=0.95\linewidth]{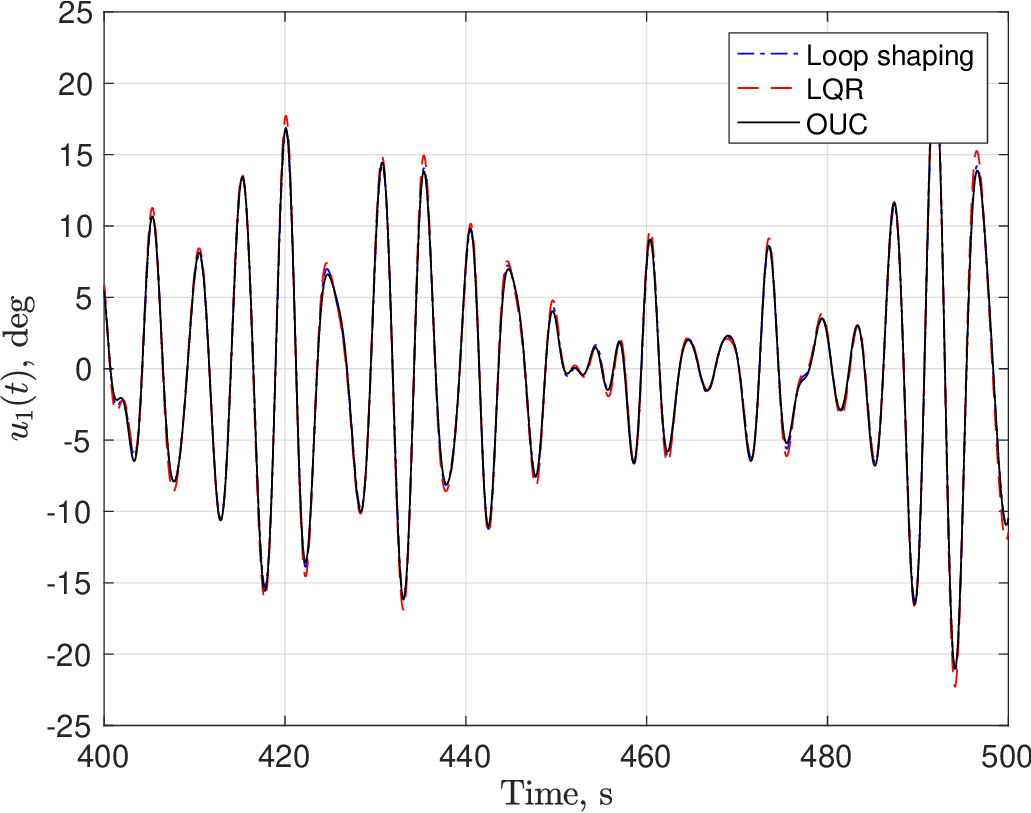} 
\end{subfigure}
\begin{subfigure}{0.49\textwidth}
\includegraphics[width=0.95\linewidth]{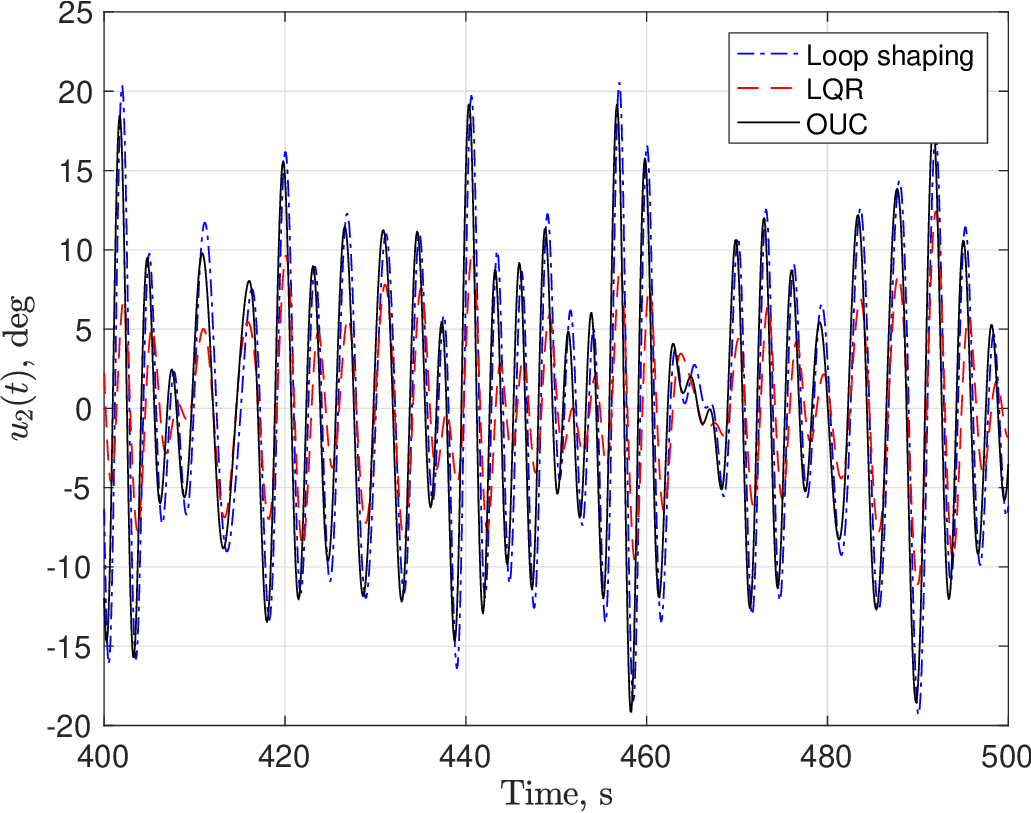}
\end{subfigure}
\caption{The result of simulation in Case 6: roll angle $e_1$, heading error $e_2$, rudder and fin angles ($u_1,u_2$).}\label{fig:case6}
\end{figure*}

{Figs.~\ref{fig:case1}--\ref{fig:case6long} the results of simulation in Cases 1-6 are presented. 
The simulation time is 500s, in Figs.~\ref{fig:case1}--\ref{fig:case6} we zoom the time window 400-500s in order to simplify viewing. Fig.~\ref{fig:case6long} shows the dynamics of roll angle for all 6 cases 
over the whole period of 500s.} As has been mentioned, we always combine the controllers with the AGC algorithm.
\begin{figure*}[h]
\centering
\begin{subfigure}{\textwidth}
\includegraphics[width=0.95\textwidth]{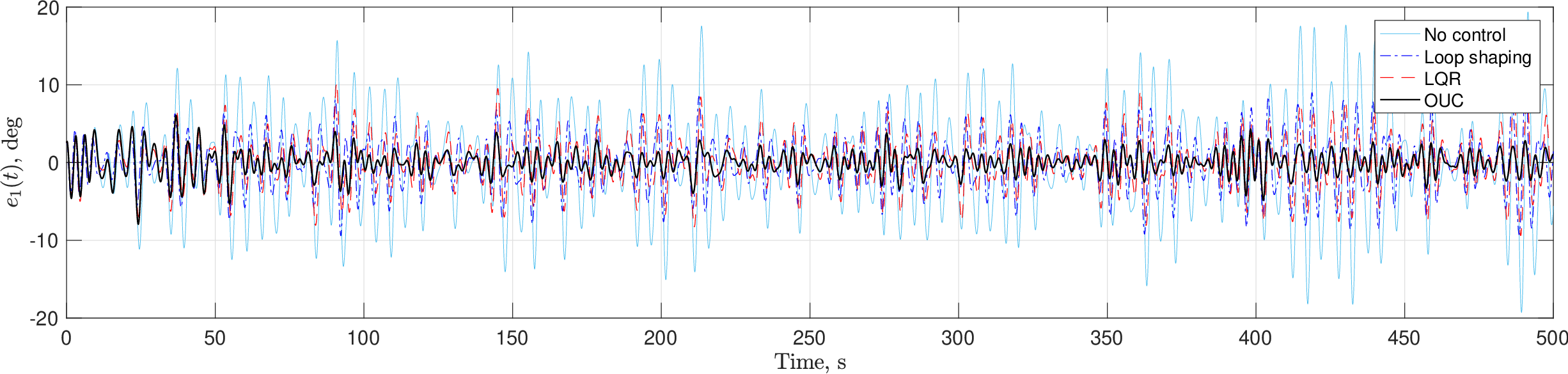}
\end{subfigure}\\[3mm]
\begin{subfigure}{\textwidth}
\includegraphics[width=0.95\textwidth]{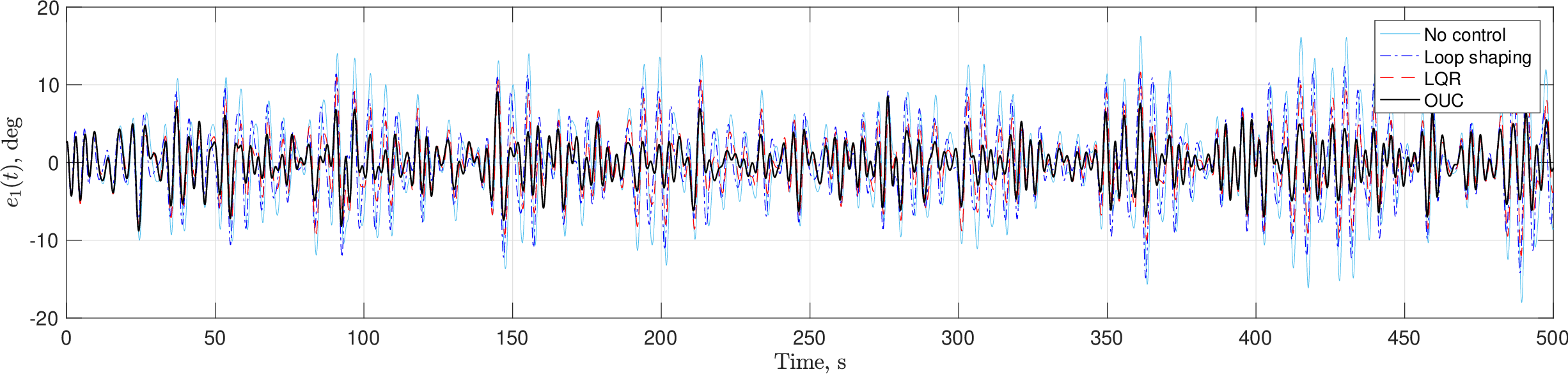}
\end{subfigure}\\[3mm]
\begin{subfigure}{\textwidth}
\includegraphics[width=0.95\textwidth]{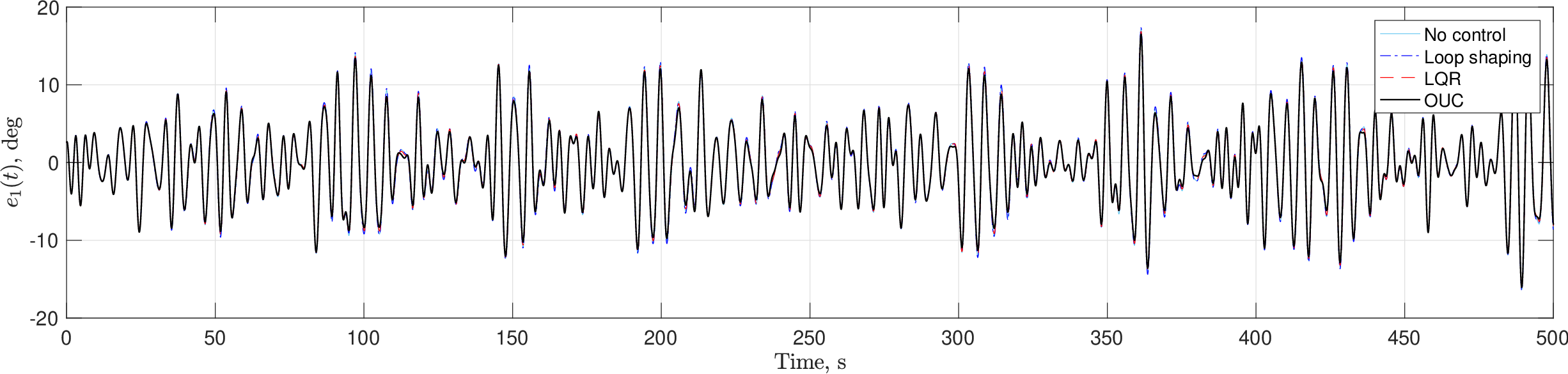}
\end{subfigure}\\[3mm]
\begin{subfigure}{\textwidth}
\includegraphics[width=0.95\textwidth]{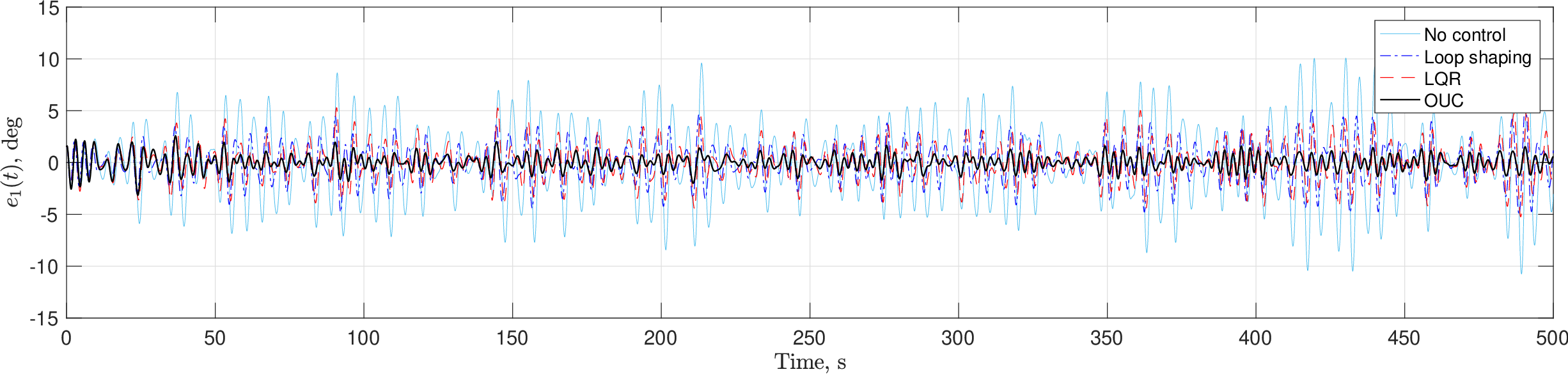}
\end{subfigure}\\[3mm]
\begin{subfigure}{\textwidth}
\includegraphics[width=0.95\textwidth]{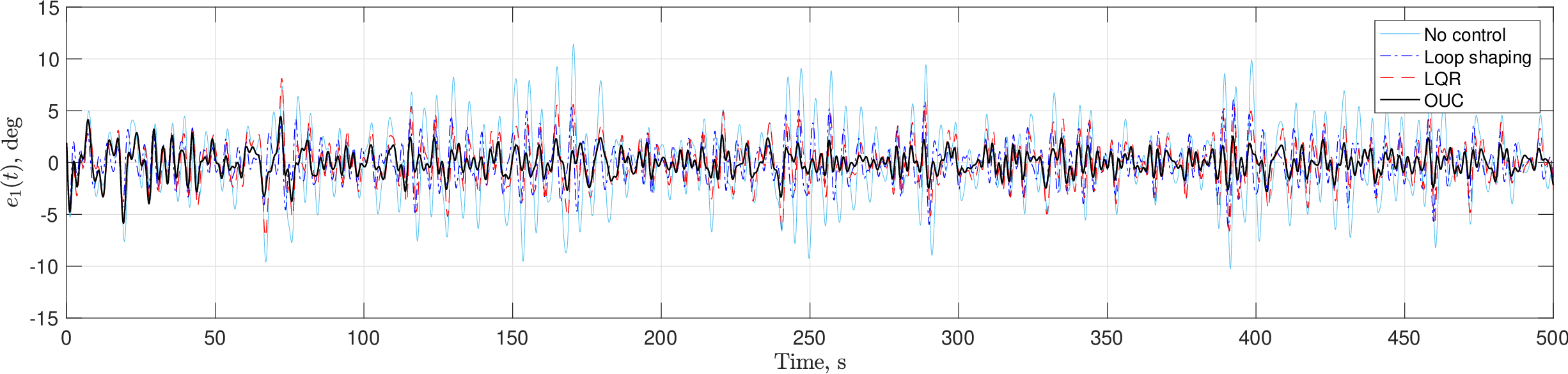}
\end{subfigure}\\[3mm]
\begin{subfigure}{\textwidth}
\includegraphics[width=0.95\textwidth]{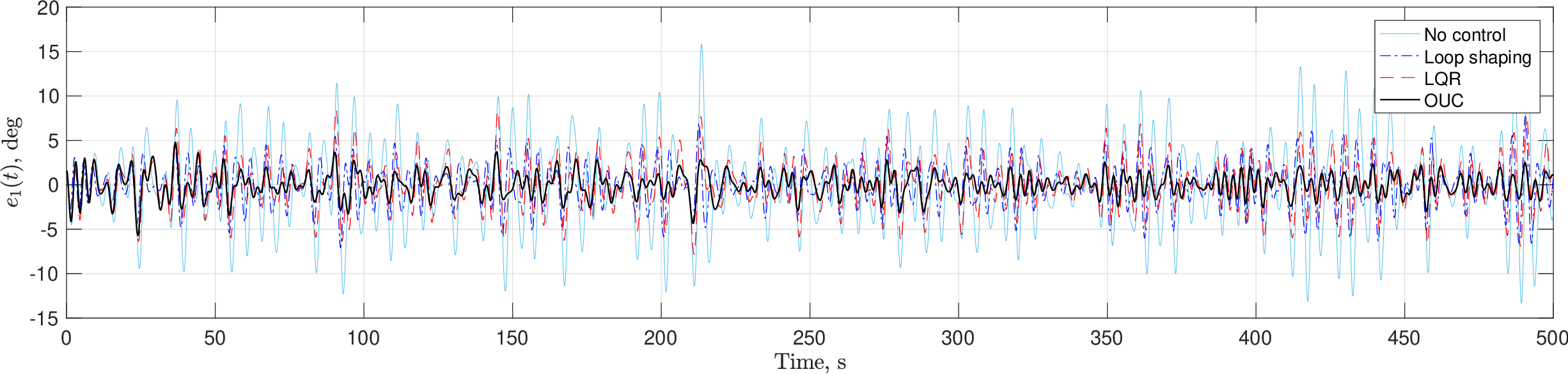}
\end{subfigure}\\[3mm]                 
\caption{Roll angle $e_1$ dynamics in Cases 1-6: full time of simulation.}\label{fig:case6long}
\end{figure*}
{The standard deviations of roll angle and heading errors are summarized in Tables~\ref{tab:stdRoll},~\ref{tab:stdYaw} respectively.} 

{
% \begin{center}
\begin{table}
\caption{STD values for roll angle \label{tab:stdRoll}}
\begin{tabularx}{0.45\textwidth}{|X|X|X|X|X|}
\hline
 Case & No \mbox{control} & OUC & LQR & Loop \mbox{Shaping}\\
%  \endfirsthead
 \hline\hline
 1 & 6.2048  &  1.4402  &  3.0443  &  3.2640 \\ 
 \hline
 2 & 5.6845  &  2.6524  &  3.7840  &  4.6454 \\
 \hline
 3 & 5.0869  &  4.8803  &  4.9800  &  5.1226 \\
 \hline
 4 & 3.5884  &   0.7313  &  1.7237  &  1.7790 \\
 \hline
 5 & 3.6071  &  1.5213  &  2.0867  &  1.7539 \\
 \hline
 6 & 4.7570  &  1.8683  &  2.5839  &  2.3128 \\
 \hline
\end{tabularx}
\end{table}
% \end{center}
}
{
% \begin{center}
\begin{table}
\caption{STD values for heading error \label{tab:stdYaw}}
\begin{tabularx}{0.45\textwidth}{|X|X|X|X|X|}
\hline
 Case & No \mbox{control} & OUC & LQR & Loop \mbox{Shaping}\\
%  \endfirsthead
 \hline\hline
 1 & 1.0746 &  1.0040  &  1.0202 &  1.0268 \\ 
 \hline
 2 & 1.0788 &  1.0532  &  1.0617 &  1.0977 \\
 \hline
 3 & 1.0169 &  1.0244  &  1.0186 &  1.0196 \\
 \hline
 4 & 0.6076 &  0.5644  &  0.5747 &  0.5780 \\
 \hline
 5 & 0.5894 &   0.5541  &  0.5574   & 0.5643 \\
 \hline
 6 & 0.7959  &  0.7501  &  0.7551  &  0.7638\\
 \hline
\end{tabularx}
\end{table}
% \end{center}
}
{Notice that in Case 3 (the speed is too small) all controllers are equally inefficient since the performance of rudders and fins is very limited. In all other cases, OUC demonstrates smaller average roll than the remaining controllers providing a comparable yaw error. Notice that this also holds in Case 6, although the actual peak frequency of the signal is different from the nominal frequencies of the wave disturbances $\omega_1,\omega_2$. The main disadvantage of the LQR is the lack of knowledge about the structure of the disturbance signal. The LQR controller shows better performance than loop shaping in Cases 1-4, because it 
efficiently damps the ship roll natural (resonance) frequency of the vessel~\citep{Perez2012129} ($\approx 1.1$s),
which is very close to the peak frequency of the wave disturbance. In Cases 5 and 6, the significant energy of the spectrum is beyond the vicinity of resonance frequency, for this reason, the performance of the simple loop-shaping controller is better.}

\subsection{Choice of the polynomial $\rho(s)$ and its influence on the closed-loop system}\label{subsec:choice}

Mathematically, every choice of the scalar polynomial $\rho(s)$ is feasible (provided that it has a sufficiently large degree and is Hurwitz). In practice, its choice influences the characteristics of the closed-loop system since the closed-loop transfer functions from the disturbance to the, respectively, control and the output are
\[
\begin{gathered}
W_{ud}(s)=\frac{r(s)\Delta(s)}{\rho(s)},\\ W_{yd}(s)=W_{yu}^0(s)W_{ud}(s)+W_{yd}^0(s).
\end{gathered}
\]
Here $W_{yu}^0=D+CA_{s}^{-1}B$ and $W_{yd}^0=G+CA_{s}^{-1}E$ are the stable open-loop transfer functions, independent of the choice of the controller. Hence, the choice of the $\rho(s)$ determines, first, the stable \emph{poles} of the closed-loop system and, second, its frequency-domain characteristics. 
The interpolation conditions provide optimal attenuation of the spectrum in a small vicinity of the nominal frequencies $\omega_1,\ldots,\omega_p$. As stated in Remark~\ref{rem.bandpass}, attenuation of the disturbance frequencies beyond this vicinity is most critical for the overall system performance.

The influence of $\rho(s)$ on the closed-loop system is illustrated in Fig.~\ref{fig:bode}. The latter figure shows the magnitude Bode plots of the functions $W_{e_1d_{\vp}}$ and $W_{u_1d_{\vp}}$ (the influence of roll disturbance on the roll angle and rudder angle), corresponding to the polynomial $\rho(\tau s)$ with $\rho(s)$ from~\eqref{eq.rho} and $\tau=0.8,1,1.2$. 
% Notice that the main energy of $d_{\vp}$ signal, according to the JONSWAP models, is concentrated in the frequency band $[0,2.5]rad/s$. 
Here $\tau$ plays the role of the system's sensitivity. The empirical observation shows that, as one decreases $\tau$ (the system becomes ``slower'', or less sensitive), the performance with respect to roll angle deteriorates (in particular, the passband becomes wider), whereas increase in $\tau$ leads to better stabilization. At the same time, large values of $\tau$ correspond to excessive use of the actuators, whereas smaller values of $\tau$ make their dynamics more smooth.
\begin{figure*}[h]
\centering
\begin{subfigure}{0.49\textwidth}
\includegraphics[width=0.95\linewidth]{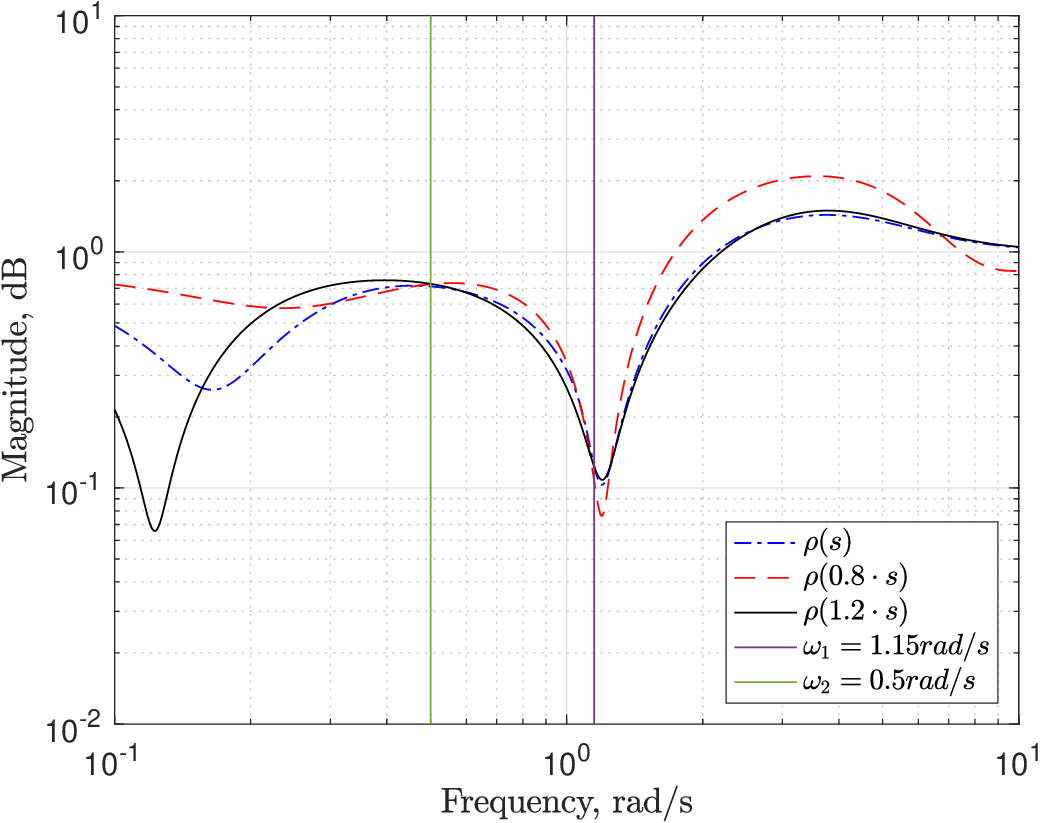}
\caption{$W_{yd}(s)$ from $d_{\varphi}$ to $e_{\phi}$}
\end{subfigure}
\begin{subfigure}{0.49\textwidth}
\includegraphics[width=0.95\linewidth]{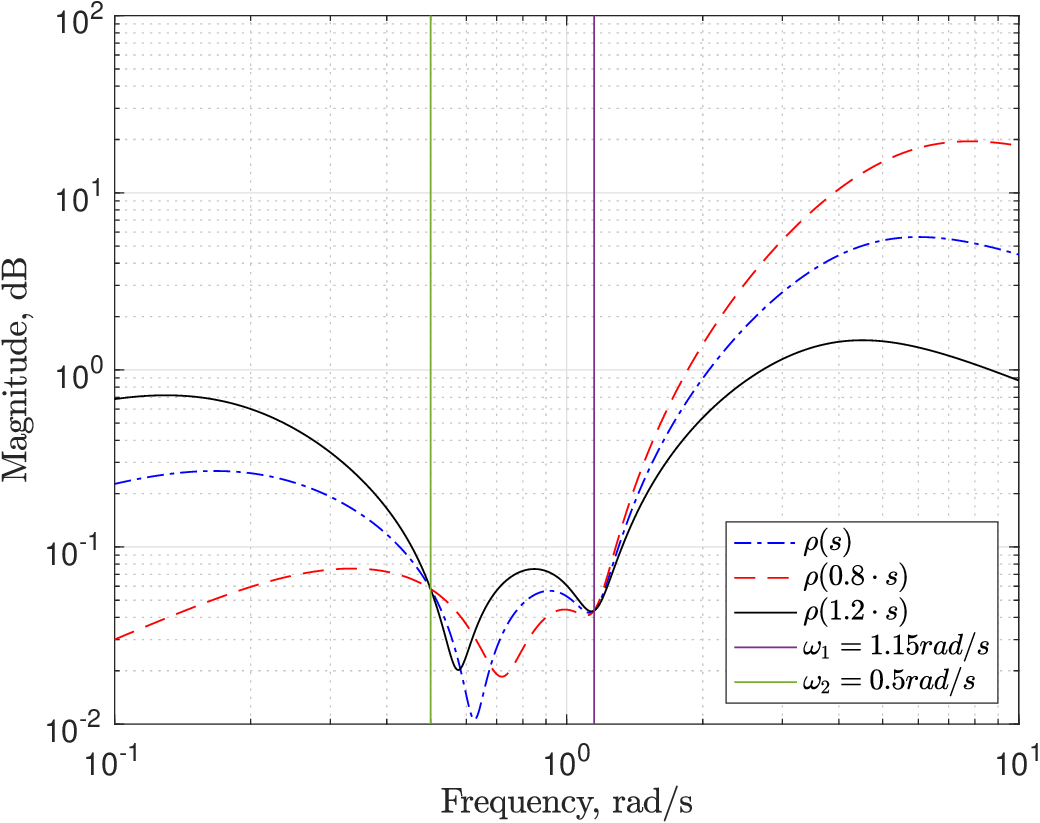}
\caption{$W_{ud}(s)$ from $d_{\varphi}$ to $u_1$}
\end{subfigure}
\caption{Bode magnitude diagrams of the closed-loop transfer functions, corresponding to $\rho(\tau s)$, with $\tau=0.8,1,1.2$.}\label{fig:bode}
\end{figure*}

The ``optimal'' assignment of closed-loop system's poles and shaping the transfer functions of the closed-loop system are long-standing problems in control theory~\citep{franklin:1981}. The denominator $\rho(s)$ of the closed-loop transfer function $W_{ud}(s)$ can be decomposed into the product
\[
\rho(s)=\sum_j(1+T_js)\sum_j(s^2+2\zeta_j\omega_{nj}s+\omega_{nj}^2),
\]
where $\omega_{nj}$ are the so-called natural frequencies, $\zeta_j$ are called damping ratios and $T_j$ are called time constants~\cite[Chapter~6]{franklin:1981}. Unlike the classical situation, the numerator of the transfer function depends on $\rho(s)$ due to the interpolation constraints~\eqref{eq:16}; also, the order of this transfer function ($\deg\rho$) has to be sufficiently large in order to satisfy the interpolation constraints. In spite of this, the standard recommendations on the pole placement~\citep{franklin:1981}, applied to the choice of $\rho(s)$, give a satisfactory result, as demonstrated by simulations in Section~5.4. One of these recommendations is to place the poles of the closed-loop system by using a standard LQR procedure. Since we are comparing our controller's behaviors with a specified LQR controller, it is natural to assign the corresponding closed-loop system's poles (zeros of $\Delta_{LQR}$) to be the roots of $\rho(s)$. To attenuate frequencies beyond $\omega_1$, $\omega_2$ and $\omega_3=0$, we also include the two Butterworth polynomials 
$(s^2+\sqrt{2}\omega_1+\omega_1^2)$, $(s^2+\sqrt{2}\omega_2+\omega_2^2)$ and an additional multiplier $1+Ts$.
The value of $T=0.5$ is chosen in order to provide the attenuation of the disturbance in the frequency band
$[0,2.2]$rad/s, which contains 93\% of the roll disturbance energy. Also, we renormalize $\rho(s)$ to set its leading coefficient to $1$. This leads us to the polynomial $\rho(s)$ from~\eqref{eq.rho}.}

\color{black}

\section{Conclusions and future work}

In this paper, we offer a novel approach to the design of the roll stabilization system for marine vessels,
based on the idea of optimal universal controllers (OUC). Unlike the existing approaches, such a controller does not require the full information about the wave's spectral density, but only the knowledge of its dominant frequencies. A topic of ongoing research is to employ adaptive control methods to enable the controller's functioning in the fully uncertain environment. In particular, combining the roll stabilization controller with an estimator of the dominating encounter wave frequencies~\citep{6145613,Belleter201548}, one can adjust the coefficients of the OUC controller ``on the fly''. 

{In our simulations, the OUC has been enhanced by an automatic gain controller (AGC)~\citep{VANAMERONGEN1990679,van1987rudder} preventing saturation of actuators. It could also be used with alternative
gain scheduling techniques, e.g. time-varying gain reduction~\citep{Lauvdal1998} or, more generally, advanced
control allocation methods taking into account nonlinear dynamics of actuators~\citep{ZACCARIAN20091431,Johansen2013}. Mathematical analysis of the resulting nonlinear systems remains, however, a non-trivial problem for future research.}

{Another limitation that can be relaxed is the fixed cruising speed of the vessel (determining the point of linearization, see~\ref{app:4dof}). Our simulation (Section~5.3) shows that the OUC controller is quite robust to the change of speed, although the optimality of the cost function is no longer guaranteed.
For vessel's maneuvers at a non-constant speed, more sophisticated controllers can be designed that are based on the paradigm of gain scheduling~\citep{RughShamma2000}. Note however that the choice of a specific
procedure allowing to redesign a linear fixed-speed controller into a nonlinear controller, applicable for non-constant speed, remains a non-trivial open problem. Also, such a redesign should be applied not only to RRS system, but also to the ship's autopilot.}

%We also want to point out that in this work we have not considered the influence of the saturation in the actuators on the %performance of a motion control system focusing on the design procedure and comparison with similar linear controllers. Further %consideration of these effects is an essential part of our ongoing research; however, it is possible to easily augment the OUC with %either a heuristic algorithm such as an automatic gain controller (AGC)~\citep{VANAMERONGEN1990679,van1987rudder}, time-varying %gain reduction (TGR)~\citep{Lauvdal1998} or more sophisticated methods based on the ideas of the control allocation with nonlinear %servo-mechanism~\citep{ZACCARIAN20091431,Johansen2013}.

\appendix

{\section{Linearized 4-DoF vessel motion model}\label{app:4dof}

    	A ship in a seaway moves in 6-DoF: three translation displacements (surge, sway and heave) define 
    	the location and three angular displacements (roll, pitch and yaw) define the attitude. In traditional maneuvering problems (such as e.g. course-keeping), normally a 3-DoF model (surge-sway-yaw) is considered. However, to consider the roll stabilization problem, one needs a 4-DoF model that includes the roll motion. 
    	In this paper, we use the Christensen and Blanke model~\cite[Section 9.1.3]{fossen1994guidance}.
	 The following equations of motion are valid when the body-fixed axes correspond to the longitudinal, lateral and normal directions:
	\[
	\begin{aligned}
		m \left[\dot{u} - y^b_g \dot{r} - v r - x^b_g r^2 + z^b_g p r \right] &=& \tau^b_{surge} \\
		m \left[ \dot{v} - z^b_g \dot{p} + x^b_g \dot{r} + u r - y^b_g (r^2 + p^2) \right] &=&\tau^b_{sway} \\
		I^b_{xx} \dot{p} - m z^b_g \dot{v} + m \left[ y^b_g v p - z^b_g u r \right] &=& \tau^b_{roll}  \\
		I^b_{zz} \dot{r} + m x^b_g \dot{v} - m y^b_g \dot{u} + m \left[ x^b_g u r + y^b_g v r \right] &=& \tau^b_{yaw}
	\end{aligned}
	\]
	where m is the mass of the ship; $x^b_g$, $y^b_g$, $z^b_g$ are the coordinates of the ship's center of gravity with respect to the body frame; $I^b_{xx}$ and $I^b_{zz}$ are the corresponding diagonal components of the inertia tensor with respect to the body frame; $u$ and $v$ are the surge and sway velocities; $p$ and $r$ are the roll and yaw rates. The total vector of forces $\tau^b$ given with respect to the body frame can be decomposed as
	\[
	\tau^b = \tau^b_{hyd} + \tau^b_{hs} + \tau^b_{c} + \tau^b_{p},
	\]
	where $\tau^b_{hyd} $, $\tau^b_{hs}$, $\tau^b_{c}$ and  $\tau^b_{p}$ stand for hydrodynamic, hydrostatic, actuators (fins and rudders) and propulsion forces and moments respectively. 
	
	It is usually assumed that the propulsion forces are compensated by the hydrodynamic resistance of the ship's hull
	$\tau^b_p+\tau_{hyd}^b=0$
	and the surge acceleration is very small, that is, $\dot{u} \approx 0$ and $u \approx U$, where $U$ is the service speed of the vessel. This leads to a simplified model to the following form:
	\begin{equation}\label{nonlin}
	\dot{x} = M^{-1} f(x) + M^{-1}\tau^b_c
	\end{equation}
	where, by definition, 
	\[M=
	\begin{bmatrix}
	 m - Y_{\dot{v}} & -m z^b_g - Y_{\dot{p}} & m x^b_g - Y_{\dot{r}} & 0 & 0\\
	-m z^b_g - K_{\dot{v}} & I^b_{xx} - K_{\dot{p}} & - K_{\dot{r}} & 0 & 0\\
	m x^b_g - N_{\dot v} & - N_{\dot p} & I^b_{zz} - N_{\dot{r}} & 0 & 0 \\
	0 & 0 & 0 & 1 & 0\\
	0 & 0 & 0 & 0 & 1
	\end{bmatrix},
	\]
	\[
	x = \begin{bmatrix}
	v \\ p \\ r \\ \phi \\ \psi
	\end{bmatrix}, \quad
	\tau^b_c = \begin{bmatrix}
	\tau^{sway}_{hyd} - m u r \\ 
	\tau^{roll}_{hyd} + m z^b_g u r \\
	\tau^{yaw}_{hyd} - m x^b_g u r \\
	0 \\ 0
	\end{bmatrix},
	\]
	where $\phi$ is the roll angle, $\psi$ is the yaw angle;
	$Y_{i}$, $K_{i}$ and $N_{i}$ stand for the hydrodynamic derivative~\citep{perez2006ship} of the sway force, roll moment and yaw moment with respect to ${i}=\dot v,\dot p,\dot r$ term; $\tau^{sway}_{hyd}$, $\tau^{roll}_{hyd}$ and $\tau^{yaw}_{hyd}$ are the nonlinear hydrodynamic forces that are found from
	\[
	\begin{aligned}
	\tau^{sway}_{hyd} =& Y_{|u|v} |U|v + Y_{ur} U r + Y_{v|v|}v|v| + Y_{v|r|} v |r|\\
	+& Y_{r |v|} r |v| +
	 Y_{\phi |u v | } \phi |Uv| + Y_{\phi |ur|} \phi |Ur| \\
	 +& Y_{\phi u u} \phi U^2 ,\\
	\tau^{roll}_{hyd} =& K_{|u| v} |U| v + K_{ur} U r + K_{v|v|} v|v| + K_{v|r|}v|r| \\
	+& K_{r|v|}r|v| + 
	+ K_\phi |uv| \phi |Uv| + K_{\phi |ur|} \phi |Ur| \\
	+&  K_{\phi uu} \phi U^2 + K_{|u|p}|U|p +\\
	+& K_{p|p|} p|p| + K_p p + K_{\phi \phi \phi} \phi^3 - \rho g \nabla G Z(\phi),\\
	\tau^{yaw}_{hyd} =& N_{|u|v}|U|v + N_{|u|r}|U|r + N_{r|r|}r|r| + N_{r|v|} r|v|\\
	+& N_{\phi |uv|} \phi |Uv| 
	+ N_{\phi u |r|} \phi U |r| + N_p p + N_{p|p|}p|p|\\ 
	+& N_{|u|p}|U|p + N_{\phi u |u|} \phi U |U|,  
	\end{aligned}
	\]
	where $\rho$ is the water density, $g$ is the acceleration of free fall, $GZ(\phi)$ is the so-called roll righting arm~\citep{perez2006ship}, and $\nabla$ is is the displaced volume.
	
	Assuming that the changes of roll and yaw angles are small one can linearize the model (\ref{nonlin}) around the equilibrium point $v = 0$, $p = 0$, $r= 0$, $\phi = 0$, $\psi = 0$ and $u = U$, i.e. the vessel is moving straight with the constant speed. That bring us to the following linear model:
	\begin{equation*}%\label{eq:vessel_model}
	\begin{aligned}
	\dot{x}(t) &= A x(t) + \tilde B \tau^b_c(t)\\
	y(t) &= C x(t) + G d(t),\\
	\end{aligned}
	\end{equation*}
	where the coefficients are found from~\eqref{eq_strip} and
	\[
	\tau_c^b(t)=U^2 L_r
	\begin{bmatrix}
	 1\\ - l^r_z \\ -l^r_x
	\end{bmatrix} \delta_{rud} + 
	U^2 L_f
	\begin{bmatrix}
	- \sin\left( \xi \right) \\ 2 l^f_r \\ l^f_x  \sin\left( \xi \right)
	\end{bmatrix} \delta_{fin},
	\]
	where
	\[
	L_r = \frac{1}{2}\rho A_r  \frac{\partial C_L^r (\delta^e_{rud})}{\partial \delta^e_{rud}},
	\]
	\[
	L_f = \frac{1}{2}\rho A_f  \frac{\partial C_L^f (\delta^e_{fin})}{\partial \delta^e_{fin}},
	\]
	 are the resulting hydrodynamic forces induced on the rudder and the fins respectively; $l^r_x,l^r_y$ are respectively the longitudinal and vertical distances from the  center of gravity(CG) to the rudder, $\xi$ is the fins tilt angle defined in the aft view, $l^f_r$ is the fin roll arm, $l^f_x$ is a longitudinal distance from the CG to the fin's center of pressure;
	$C_L^r (\delta^e_{rud})$ and $C_L^f (\delta^e_{fin})$ are the lift characteristics for the rudder and fins respectively,
	%for the  effective angle of attack of the rudder $\delta^e_{rud}$ and fins $\delta^e_{fin}$ respectively; 
	$A_r$ and $A_f$ stand for to the rudder and fins areas. 
 
	%\begin{strip}
	\begin{table*}
	\begin{minipage}{0.95\textwidth}
	\begin{equation}\label{eq_strip}
	\begin{gathered}
	A = M^{-1} \left. \frac{\partial f(x)}{\partial x} \right|_{x = 0}, \
	\tilde B = M^{-1},\,
	C = \begin{bmatrix}
	0 & 0 & 0 & 1 & 0 \\
	0 & 0 & 0 & 0 & 1 
	\end{bmatrix} , \
	G = \begin{bmatrix}
	1 & 0 \\ 0 & 1
	\end{bmatrix}\\
	\left. \frac{\partial f(x)}{\partial x} \right|_{x = 0} = 
	\begin{bmatrix}
	Y_{|u|v}|U| & 0 & (Y_ur - m)U & Y_{\phi u u} U^2 & 0\\
	K_{|u|v}|U|& K_p + K_{|u|p}|U| & (K_{ur} + mz^b_g)U & K_{\phi u u} U^2 - \rho g \nabla GMt & 0\\
	N_{|u|v} |U| & N_p + N_{|u|p}|U| & N_{|u|r} |U| - m x_g U & 
	N_{\phi u |u|} U |U| & 0 \\
	0 & 1 & 0 & 0 & 0 \\
	0 & 0 & 1 & 0 & 0
	\end{bmatrix}
	\end{gathered}
	\end{equation}
	%\end{strip}
	\end{minipage}
	\medskip\hrule
	\end{table*}
}

\section{Transfer matrices of the ship-autopilot system}\label{app:Model}
In this section we are going to present the transformation procedure on how to obtain the models in equations~\eqref{eq:rel23}  based on
the general dynamics of the vessel described by the transfer function
\begin{align*}
\varphi(t) =& W_{\varphi r}\ddt \delta_{rud}(t) + W_{\varphi f}\ddt \delta_{fin}(t),\\
\psi(t) =& W_{\psi r}\ddt \delta_{rud}(t) + W_{\psi f}\ddt \delta_{fin}(t).
\end{align*}
The observed outputs of the system are
\[
e_{\psi}(t) = \psi(t) + d_{\psi}(t) - \bar{\psi}, \quad e_{\varphi} = \varphi(t) + d_\varphi(t),
\]
where $\bar\psi$ is the heading setpoint. We introduce the two control inputs as follows
\[
\begin{aligned}
u_1&=\delta_{rud}(t)-W_{AP} \ddt e_{\psi}(t),\\
u_2&=\delta_{fin}(t),
\end{aligned}
\]
where $W_{AP}$ is the autopilot's transfer function, stabilizing the vessel's yaw motion. Putting the equations together, one arrives at the following
\[\begin{aligned}
	\begin{bmatrix}
		1 & - W_{\varphi r} W_{ap} \\
		0 & 1 - W_{\psi r} W_{ap}
	\end{bmatrix}
	\begin{bmatrix}
		e_\varphi \\
		e_\psi
	\end{bmatrix}
	= 	
	\begin{bmatrix}
		W_{\varphi r} & W_{\varphi f}\\
		W_{\psi r} & W_{\psi f}
	\end{bmatrix}	
	\begin{bmatrix}
	u_1 \\ u_2
	\end{bmatrix}
	+\\+	
	\begin{bmatrix}
		0 & 1 & 0 \\
	   -1 & 0 & 1
	\end{bmatrix}	
	\begin{bmatrix}
		\bar{\psi}\\
		d_\varphi \\
		d_\psi
	\end{bmatrix}.
\end{aligned}\]

Assuming that the autopilot stabilizes the yaw loop i.e. $1 - W_{\psi r} W_{ap} \neq 0$ this yields
\begin{align*}\label{eq:rel3}
\begin{bmatrix}
e_\varphi \\
e_\psi
\end{bmatrix}
&=
\begin{bmatrix}
W_{\varphi u_1}^0 & W_{\varphi u_2}^0\\
W_{\psi u_1}^0 & W_{\psi u_2}^0
\end{bmatrix}
\begin{bmatrix}
u_1 \\ u_2
\end{bmatrix}
+\\
&+
\begin{bmatrix}
W_{\varphi \bar{\psi}}^0&1 & W_{\varphi d_\psi}^0 \\
W_{\psi \bar{\psi}}^0 &0 & W_{\psi d_\psi}^0
\end{bmatrix}
\begin{bmatrix}
\bar{\psi}\\
d_\varphi \\
d_\psi
\end{bmatrix},
\end{align*}
where
\[
\begin{gathered}
W_{\varphi u_1}^0 = (1-W_{\psi r} W_{ap})^{-1} W_{\varphi r},\\
W_{\psi u_1}^0  = (1-W_{\psi r} W_{ap})^{-1} W_{\psi r},\\
W_{\varphi u_2}^0 = W_{\varphi f} + (1-W_{\psi r} W_{ap})^{-1} W_{\varphi r} W_{ap} W_{\psi f},\\
W_{\psi u_2}^0  = (1-W_{\psi r} W_{ap})^{-1} W_{\psi f},\\
-W_{\varphi \bar{\psi}}^0 = W_{\varphi d_{\psi}}^0  = (1-W_{\psi r} W_{ap})^{-1} W_{\varphi r} W_{ap},\\
-W_{\psi \bar{\psi}}^0 = W_{\psi d_{\psi}}^0  = (1-W_{\psi r} W_{ap})^{-1}.
\end{gathered}
\]

Recalling that
\[
y = 
\begin{bmatrix}
e_\varphi \\
e_\psi
\end{bmatrix}, \quad
u =
\begin{bmatrix}
u_1 \\
u_2
\end{bmatrix}, \quad
d = 
\begin{bmatrix}
\bar{\psi}\\
d_\varphi \\
d_\psi
\end{bmatrix},
\]
the transfer matrices from $u$ and $d$ respectively to $y$ are given by
\[
W_{yu}^0 = 
\begin{bmatrix}
W_{\varphi u_1}^0 & W_{\varphi u_2}^0\\
W_{\psi u_1}^0 & W_{\psi u_2}^0
\end{bmatrix},
  W_{yd}^0 = 
\begin{bmatrix}
W_{\varphi \bar{\psi}}^0&1 & W_{\varphi d_\psi}^0 \\
W_{\psi \bar{\psi}}^0&0 & W_{\psi d_\psi}^0
\end{bmatrix}.
\]

%\section*{References}

\bibliographystyle{elsarticle-harv}
\bibliography{ref}

\begin{thebibliography}{47}
\expandafter\ifx\csname natexlab\endcsname\relax\def\natexlab#1{#1}\fi
\expandafter\ifx\csname url\endcsname\relax
  \def\url#1{\texttt{#1}}\fi
\expandafter\ifx\csname urlprefix\endcsname\relax\def\urlprefix{URL }\fi

\bibitem[{Anderson and Moore(1990)}]{Anderson:1990:OCL:79089}
Anderson, B. D.~O., Moore, J.~B., 1990. Optimal Control: Linear Quadratic
  Methods. Prentice-Hall, Inc., Upper Saddle River, NJ, USA.

\bibitem[{Balloch(1998)}]{Balloch1998}
Balloch, R., 1998. Attitude sensors for {DP}. In: Proc. Dynamic Positioning
  Conf. Houston, TX, USA.

\bibitem[{Belleter et~al.(2015)Belleter, Galeazzi, and Fossen}]{Belleter201548}
Belleter, D.~J., Galeazzi, R., Fossen, T.~I., 2015. Experimental verification
  of a global exponential stable nonlinear wave encounter frequency estimator.
  Ocean Engineering 97, 48 -- 56.

\bibitem[{Blanke et~al.(2000)Blanke, Adrian, Larsen, and
  Bentsen}]{blanke2000rudder}
Blanke, M., Adrian, J., Larsen, K.-E., Bentsen, J., 2000. Rudder roll damping
  in coastal region sea conditions. In: Proc. 5th IFAC Conference on
  Manoeuvring and Control of Marine Craft (MCMC). Aalborg, Denmark, pp. 30--44.

\bibitem[{Blanke and Christensen(1993)}]{blanke1993rudder}
Blanke, M., Christensen, A.~C., 1993. Rudder-roll damping autopilot robustness
  to sway-yaw-roll couplings. 10th Ship Control Systems Symposium, Ottawa 25-29
  Oct. 1993.

\bibitem[{Bobtsov et~al.(2012)Bobtsov, Efimov, Pyrkin, and Zolghadri}]{6145613}
Bobtsov, A.~A., Efimov, D., Pyrkin, A.~A., Zolghadri, A., Sept 2012. Switched
  algorithm for frequency estimation with noise rejection. IEEE Transactions on
  Automatic Control 57~(9), 2400--2404.

\bibitem[{Carley and Duberley(1972)}]{ruddFinInterraction}
Carley, J., Duberley, A., 1972. Design considerations for optimum ship motion
  control. In: Proceedings of 3rd Ship Control Systems Symposium, Volume-C.
  Bath, UK.

\bibitem[{Carley(1975)}]{Carley1975}
Carley, J.~B., 1975. Feasibility study of steering and stabilising by rudder.
  In: Proc. 4th Ship control system symposium. The Hague, The Netherlands.

\bibitem[{Cowley and Lambert(1972)}]{CowleyLambert1972}
Cowley, W., Lambert, T., 1972. The use of the rudder as a roll stabiliser. In:
  Proc. 3rd Ship Control Syst. Symp. Bath, UK.

\bibitem[{Crossland(2003)}]{Crossland2003423}
Crossland, P., 2003. The effect of roll-stabilisation controllers on warship
  operational performance. Control Engineering Practice 11~(4), 423 -- 431.

\bibitem[{Fedele and Ferrise(2012)}]{Fedele2012}
Fedele, G., Ferrise, A., 2012. Non adaptive second-order generalized integrator
  for identification of a biased sinusoidal signal. IEEE Trans. Autom. Control
  57~(7), 1838--1842.

\bibitem[{Fossen(1994)}]{fossen1994guidance}
Fossen, T., 1994. Guidance and control of ocean vehicles. Wiley.

\bibitem[{Fossen and Perez(2004)}]{mss_toolbox}
Fossen, T.~I., Perez, T., 2004. Marine systems simulator ({MSS}).
  \url{https://github.com/cybergalactic/MSS}.

\bibitem[{Franklin et~al.(1981)Franklin, Powell, and
  Emami-Naemi}]{franklin:1981}
Franklin, G., Powell, J., Emami-Naemi, A., 1981. Feedback Control of Dynamic
  Systems. Advances in Industrial Control. Addison-Wesley, Reading, MA.

\bibitem[{Goodwin et~al.(2000)Goodwin, Perez, Seron, and
  Tzeng}]{GoodwinPerez914671}
Goodwin, G.~C., Perez, T., Seron, M., Tzeng, C.~Y., 2000. On fundamental
  limitations for rudder roll stabilization of ships. In: Proc. of IEEE Conf.
  Decision and Control (CDC). Vol.~5. pp. 4705--4710.

\bibitem[{Hearns and Blanke(1998)}]{hearns98}
Hearns, G., Blanke, M., 1998. Quantitative analysis and design of a rudder roll
  damping controller. In: Proc. of IFAC Conference on Control Applications in
  Marine Systems (CAMS). Fukuoka, Japan, pp. 115--120.

\bibitem[{Hinostroza et~al.(2015)Hinostroza, Luo, and Soares}]{SOARES2015126}
Hinostroza, M., Luo, W., Soares, C.~G., 2015. Robust fin control for ship roll
  stabilization based on {$L_2$}-gain design. Ocean Engineering 94, 126 -- 131.

\bibitem[{Horowitz and Sidi(1978)}]{doi:10.1080/00207177808922376}
Horowitz, I., Sidi, M., 1978. Optimum synthesis of non-minimum phase feedback
  systems with plant uncertainty. International Journal of Control 27~(3),
  361--386.

\bibitem[{Hou(2012)}]{Hou2012}
Hou, M., 2012. Parameter identification of sinusoids. IEEE Trans. Autom.
  Control 57~(2), 467--472.

\bibitem[{Johansen and Fossen(2013)}]{Johansen2013}
Johansen, T., Fossen, T., 2013. Control allocation -- a survey. Automatica
  49~(5), 1087--1103.

\bibitem[{Johansen et~al.(2008)Johansen, Fuglseth, T{\o}ndel, and
  Fossen}]{Johansen2008}
Johansen, T., Fuglseth, T., T{\o}ndel, P., Fossen, T., 2008. Optimal
  constrained control allocation in marine surface vessels with rudders.
  Control Engineering Practice 16, 457--464.

\bibitem[{Kapitanyuk et~al.(2016)Kapitanyuk, Proskurnikov, and Cao}]{cams}
Kapitanyuk, Y.~A., Proskurnikov, A.~V., Cao, M., 2016. Optimal controllers for
  rudder roll damping with an autopilot in the loop. IFAC-PapersOnLine 49~(23),
  562 -- 567, 10th IFAC Conference on Control Applications in Marine
  SystemsCAMS 2016.

\bibitem[{Lauvdal and Fossen(1997)}]{Lauvdal95nonlinearnon-minimum}
Lauvdal, T., Fossen, T., 1997. Nonlinear non-minimum phase rudder-roll damping
  system for ships using sliding mode control. In: Proc. Europ. Control
  Conference. Brussels, Belgium, pp. 1689--1694.

\bibitem[{Lauvdal and Fossen(1998)}]{Lauvdal1998}
Lauvdal, T., Fossen, T.~I., 1998. Rudder roll stabilization of ships subject to
  input rate saturation using a gain scheduled control law. IFAC Proceedings
  Volumes 31~(30), 111 -- 116, iFAC Conference on Control Applications in
  Marine Systems (CAMS '98), Fukuoka, Japan, 27-30 October.

\bibitem[{Lindquist and Yakubovich(1997)}]{587333}
Lindquist, A., Yakubovich, V.~A., 1997. Optimal damping of forced oscillations
  in discrete-time systems. IEEE Transactions on Automatic Control 42~(6),
  786--802.

\bibitem[{Lindquist and Yakubovich(1999)}]{788535}
Lindquist, A., Yakubovich, V.~A., 1999. Universal regulators for optimal
  tracking in discrete-time systems affected by harmonic disturbances. IEEE
  Transactions on Automatic Control 44~(9), 1688--1704.

\bibitem[{Liu et~al.(2016)Liu, Jin, Grimble, and Katebi}]{GRIMBLE}
Liu, Z., Jin, H., Grimble, M.~J., Katebi, R., 2016. Ship forward speed loss
  minimization using nonlinear course keeping and roll motion controllers.
  Ocean Engineering 113, 201 -- 207.

\bibitem[{Lloyd(1975)}]{Lloyd1975}
Lloyd, A., 1975. Roll stabilisation by rudder. In: Proc. 4th Ship control
  system symposium. The Hague, The Netherlands.

\bibitem[{Longuet-Higgins(1963)}]{irregular_waves}
Longuet-Higgins, M.~S., 1963. The effect of non-linearities on statistical
  distributions in the theory of sea waves. Journal of Fluid Mechanics 17~(3),
  459?480.

\bibitem[{Marzouk and Nayfeh(2009)}]{rollTanks}
Marzouk, O.~A., Nayfeh, A.~H., 2009. Control of ship roll using passive and
  active anti-roll tanks. Ocean Engineering 36~(9), 661 -- 671.

\bibitem[{Nicolau et~al.(2005)Nicolau, Miholc{\u a}, Aiordachioaie, and
  Ceang{\u a}}]{Nicolau2005}
Nicolau, V., Miholc{\u a}, C., Aiordachioaie, D., Ceang{\u a}, E., 2005. {QFT}
  autopilot design for robust control of ship course-keeping and
  course-changing problems. Control Engineering and Applied Informatics 7~(1),
  44--56.

\bibitem[{Perez(2006)}]{perez2006ship}
Perez, T., 2006. Ship Motion Control: Course Keeping and Roll Stabilisation
  Using Rudder and Fins. Advances in Industrial Control. Springer London.

\bibitem[{Perez and Blanke(2012)}]{Perez2012129}
Perez, T., Blanke, M., 2012. Ship roll damping control. Annual Reviews in
  Control 36~(1), 129 -- 147.

\bibitem[{Proskurnikov and Yakubovich(2003{\natexlab{a}})}]{ProYak:03b}
Proskurnikov, A., Yakubovich, V., 2003{\natexlab{a}}. Approximate solution to
  the problem of the invariance of a control system. Doklady Mathematics
  68~(2), 308--312.

\bibitem[{Proskurnikov and Yakubovich(2003{\natexlab{b}})}]{ProYak:03a}
Proskurnikov, A., Yakubovich, V., 2003{\natexlab{b}}. The problem of the
  invariance of a control system. Doklady Mathematics 67~(2), 291--295.

\bibitem[{Proskurnikov(2015)}]{Proskurnikov2015557}
Proskurnikov, A.~V., 2015. Universal controllers of {V.A. Yakubovich}: a
  systematic approach to {LQR} problems with uncertain external signals.
  IFAC-PapersOnLine 48~(11), 557 -- 562, in {P}roc. of 1st IFAC Conference on
  Modelling, Identification and Control of Nonlinear Systems, 2015 St.
  Petersburg, Russia, 24-26 June 2015.

\bibitem[{Proskurnikov and Yakubovich(2006)}]{Proskurnikov2006}
Proskurnikov, A.~V., Yakubovich, V.~A., 2006. Universal regulators for optimal
  tracking of polyharmonic signals in systems with delays. Doklady Mathematics
  73~(1), 147--151.

\bibitem[{Proskurnikov and Yakubovich(2012)}]{Proskurnikov2012}
Proskurnikov, A.~V., Yakubovich, V.~A., 2012. Universal controllers in model
  matching optimal control problems for unknown external signals. Journal of
  Computer and Systems Sciences International 51~(2), 214--227.

\bibitem[{Rugh and Shamma(2000)}]{RughShamma2000}
Rugh, W.~J., Shamma, J.~S., 2000. Research on gain scheduling. Automatica
  36~(10), 1401 -- 1425.

\bibitem[{Sharif et~al.(1995)Sharif, Roberts, and Sutton}]{Sharif1995703}
Sharif, M., Roberts, G., Sutton, R., 1995. Sea-trial experimental results of
  fin/rudder roll stabilisation. Control Engineering Practice 3~(5), 703 --
  708.

\bibitem[{Stoustrup et~al.(1994)Stoustrup, Niemann, and Blanke}]{381212}
Stoustrup, J., Niemann, H.~H., Blanke, M., Aug 1994. Roll damping by rudder
  control - a new {H}$_\infty$ approach. In: Control Applications, 1994.,
  Proceedings of the Third IEEE Conference on. pp. 839--844 vol.2.

\bibitem[{Surendran et~al.(2007)Surendran, Lee, and Kim}]{FINS2007542}
Surendran, S., Lee, S., Kim, S., 2007. Studies on an algorithm to control the
  roll motion using active fins. Ocean Engineering 34~(3), 542 -- 551.

\bibitem[{van Amerongen et~al.(1990)van Amerongen, van~der Klugt, and van
  Nauta~Lemke}]{VANAMERONGEN1990679}
van Amerongen, J., van~der Klugt, P., van Nauta~Lemke, H., 1990. Rudder roll
  stabilization for ships. Automatica 26~(4), 679 -- 690.

\bibitem[{van~der Klugt(1987)}]{van1987rudder}
van~der Klugt, P., 1987. Rudder Roll Stabilization. Ph.D. thesis, Delft
  University of Technology, The Netherlands.

\bibitem[{Veremey(2014)}]{Veremey2014}
Veremey, E., 2014. Dynamical correction of control laws for marine ships
  accurate steering. J. Marine Sci. Appl. 13, 127--133.

\bibitem[{Yakubovich(1995)}]{Yakubovich1995}
Yakubovich, V., 1995. Universal regulators in linear-quadratic optimization
  problems. In: Trends in Control: A European Perspective. Springer, London,
  pp. 53--68.

\bibitem[{Zaccarian(2009)}]{ZACCARIAN20091431}
Zaccarian, L., 2009. Dynamic allocation for input redundant control systems.
  Automatica 45~(6), 1431 -- 1438.

\end{thebibliography}

\end{document}